\documentclass[useAMS,usenatbib]{mn2e}
\usepackage{graphicx}
\usepackage{amsmath}
\usepackage{amssymb}
\usepackage{color}
\usepackage{url}
\newcommand{\xmm}{\textit{XMM-Newton}}
\newcommand{\ch}{\textit{Chandra}}

\newcommand\nodata{ ~$\cdots$~ }%

\newcommand  \acc     {\ifmmode {\rm km\,s}^{-2} \else km\,s$^{-2}$\fi}

\newcommand  \ergs     {\ifmmode {\rm ergs\,s}^{-1} \else ergs s$^{-1}$\fi}
\newcommand  \ergcms   {\ifmmode {\rm erg~cm}^{-2}\,{\rm s}^{-1}
                        \else erg~cm$^{-2}$\,s$^{-1}$\fi}
\newcommand  \ergcmsA  {\ifmmode{\rm erg\,cm}^{-2}\,{\rm s}^{-1}\,{\rm\AA}^{-1}
                        \else ergs\,cm$^{-2}$\,s$^{-1}$\,\AA$^{-1}$\fi}
\newcommand  \ergcmsHz {\ifmmode{\rm ergs\,cm}^{-2}\,{\rm s}^{-1}\,{\rm Hz}^{-1}
                        \else ergs\,cm$^{-2}$\,s$^{-1}$\,Hz$^{-1}$\fi}
\newcommand  \phcms    {\ifmmode {\rm ph\,cm}^{-2}\,{\rm s}^{-1}
                        \else ph\,cm$^{-2}$\,s$^{-1}$\fi}
\newcommand  \phcmsA   {\ifmmode {\rm ph\,cm}^{-2}\,{\rm s}^{-1}\,{\rm\AA}^{-1}
                        \else ph\,cm$^{-2}$\,s$^{-1}$\,\AA$^{-1}$\fi}

\newcommand\aj{{AJ}}%
\newcommand\araa{{ARA\&A}}%
\newcommand\apj{{ApJ}}%
\newcommand\apjl{{ApJ}}%
\newcommand\apjs{{ApJS}}%
\newcommand\aap{{A\&A}}%
\newcommand\aaps{{A\&AS}}%
\newcommand\mnras{{MNRAS}}%
\newcommand\pre{{Phys.~Rev.~E}}%
\newcommand\pasp{{PASP}}%
\title[Sizes of SNRs in the Magellanic Clouds]{On the Size Distribution of Supernova Remnants in the Magellanic Clouds}

\author[Badenes, Maoz, \& Draine] {Carles Badenes$^{1,2}$\thanks{E-mail: carles@astro.tau.ac.il},
  Dan Maoz$^{1}$, Bruce T. Draine$^{3}$\\
  $^{1}$School of Physics and Astronomy, Tel-Aviv University, Tel-Aviv 69978,
  Israel\\
  $^{2}$Benoziyo Center for Astrophysics, Weizmann Institure of Science, Rehovot 76100, Israel\\
  $^{3}$Department of Astrophysical Sciences, Princeton University, Ivy Lane, Princeton, NJ 08540\\
} \date{\today}

\begin{document}

\maketitle

\label{firstpage}

\begin{abstract}
  The physical sizes of supernova remnants (SNRs) in a number of nearby galaxies follow an approximately linear
  cumulative distribution, contrary to what is expected for decelerating shock fronts. This phenomenon has been
  variously attributed to observational selection effects, or to a majority of SNRs being in ``free expansion'', with
  shocks propagating at a constant velocity into a tenuous ambient medium.  We compile multi-wavelength observations of
  the 77 known SNRs in the Magellanic Clouds, and argue that they provide a fairly complete record of the SNe that have
  exploded over the last $\sim 20$~kyr, with most of them now in the adiabatic, Sedov phase of their expansions.  The
  roughly linear cumulative distribution of sizes (roughly uniform in a differential distribution) can be understood to
  result from the combination of the deceleration during this phase, a transition to a radiation-loss-dominated phase at
  a radius that depends on the local gas density, and a probability distribution of densities in the interstellar medium
  varying approximately as $\rho^{-1}$.  This explanation is supported by the observed powerlaw distributions, with
  index $\sim -1$, of three independent tracers of density: neutral hydrogen column density, H$\alpha$ surface
  brightness, and star-formation rate based on resolved stellar populations. In this picture, the observed cutoff at a
  radius of 30~pc in the SNR size distribution is due to a minimum in the mean ambient gas density in the regions where
  supernovae (SNe) explode. We show that M33 has a SNR size distribution very similar to that of the Magellanic Clouds,
  suggesting these features, and their explanation, may be universal. In a companion paper (Maoz \& Badenes 2010), we
  use our sample of SNRs as an effective ``SN survey'' to calculate the SN rate and delay time distribution in the
  Magellanic Clouds. The hypothesis that most SNRs are in free expansion, rather than in the Sedov phase of their
  evolution, would result in SN rates that are in strong conflict with independent measurements, and with basic stellar
  evolution theory.

\end{abstract}

\begin{keywords}
supernovae: general -- supernovae: remnants --
galaxies: individual: LMC, SMC
\end{keywords}

\section{Introduction}
The ecology of galaxies is dominated by supernova (SN) explosions, which inject energy and enriched material into
the interstellar medium and trigger the formation of the next generations of stars. Many fundamental aspects of SNe
are still poorly understood, both for the core-collapse (CC) SN explosions that are thought to end the lives of massive
($\ga 8M_\odot$) stars and for the type-Ia SNe (SNe~Ia) that are believed to be the thermonuclear combustions of CO
white dwarfs (WDs) that approach the Chandrasekhar mass. Nevertheless, the two flavors of SNe deposit a similar amount
($\sim10^{51}$ erg) of kinetic energy into the surrounding medium, leaving behind supernova remnants (SNRs) that remain visible
for thousands of years. In recent times, the study of young SNRs at X-ray wavelengths has emerged as a new way to
explore the physics of CC and Type Ia SN explosions \citep[see][and references therein]{badenes10:PNAS_review}, but most
known SNRs are too old to provide much information about the specific events that originated them \citep[see discussions
in][]{rakowski05:G337,badenes09:SNRs_LMC}. In spite of this, much can be learned by studying the entire population of
SNRs, young and old, in nearby galaxies. Because the timescales for SNR evolution are short compared to most of the
processes that affect the structure of galaxies, SNR catalogues provide a clean record of the environments where SNe
explode, which can be used to put constraints on the properties of their progenitors
\citep[e.g.][]{badenes09:SNRs_LMC}. Moreover, by considering the properties of the entire population of SNRs in a galaxy
together with the bulk properties of the gas they are expanding into, we can gain insights into the evolutionary phases
of SNRs \citep{woltjer72:SNR-review}, the structure of galaxies on scales comparable to the average SNR size
\citep{cox05:ISM}, and the cycles of matter and energy in the interstellar medium \citep{ferriere01:ISM}.

In this paper and in a companion publication (Maoz \& Badenes 2010, henceforth Paper~II), we use the SNR population in the
Magellanic Clouds to explore some of these issues. The Magellanic Clouds (MCs) have the advantage of being close enough
to study key aspects of their global structure in great detail, and they also harbor a large and extensively observed
population of SNRs. Thus, they are the optimal setting to study the interplay between local density, star formation, SN
explosions, and SNR evolution on a galactic scale. Our ultimate goal, and the focus of Paper~II, is to derive the SN rate
and delay time distribution (i.e., the SN rate as a function of time following a brief burst of star formation) in the
Magellanic Clouds. However, this cannot be done without understanding first the relationship between the lifetime of
SNRs and the properties of their local environments. This is the subject of the present work.

The evolution of SNRs has been the subject of many theoretical studies
\citep[e.g.,][]{woltjer72:SNR-review,chevalier82:selfsimilar,cioffi88:Radiative_SNRs,blondin98:Radiative_SNRs,truelove99:adiabatic-SNRs}.
Because accurate ages are only known for a handful of young, often historical objects, any observational tests of these
theoretical models must rely on SNR size as a proxy for age. Given that SNRs of equal ages will have different sizes if
they expand in different media, this necessarily brings the role of local density into the picture. Previous works on
the distribution of SNR sizes initially focused on the Milky Way and the MCs
\citep[e.g.][]{mathewson84:SNR-Magellanic-Clouds,green84:SNR_Statistics,hughes84:SNR_N_D,berkhuijsen87:SNRs_N_D,chu88:LMC_SNRs_environments},
but more recent efforts have also explored other galaxies in the Local Group, including M31
\citep{magnier97:M31_SNRs_ROSAT}, M33 \citep{long10:M33_SNRs}, and M83 \citep{dopita10:M83_SNRs}. With few exceptions,
these studies gave little consideration to the bulk properties of the gas in the galaxies hosting the SNRs. In many
cases, their samples were also affected by issues of completeness and biases from working at a single wavelength. Not
surprisingly, these efforts have failed to produce a unified, physically motivated picture of the evolution of SNRs in
the interstellar medium. Here, we propose a first approximation to the problem in the context of the Magellanic Clouds.

This paper is organized as follows. In \S~\ref{catalog}, we present a compilation of multi-wavelength observations for
the 77 known SNRs in the Magellanic Clouds, and we argue that it provides a fairly complete record of all the SNe that
have exploded over the last $\sim 20$~kyr. In \S~\ref{distribution}, we examine the size distribution of SNRs in both
galaxies, and we find that the cumulative distribution is close to linear (i.e., the differential is close to uniform),
within the uncertainties associated with the relatively small number of objects, up to a marked cutoff at a physical
radius of $\sim$ 30 pc. We also show that the SNR size distribution in M33 has very similar properties, suggesting that
these features might be widespread. In \S~\ref{physics}, we propose a physical model to explain this distribution, based
on the assumption that most objects are in the Sedov (adiabatic) stage of their evolution, and that they rapidly fade
away once they transition to the radiative stage, at an age that depends on the local density. Under these conditions,
the uniform distribution of SNR sizes requires that the gas density in the Clouds have a probability distribution
described by a power law with an index of $-$1 (i.e., $\delta P / \delta \rho \sim \rho^{-1}$). In
\S~\ref{densityestimates}, we test this requirement by examining the distribution of three independent density tracers
in the Clouds: neutral hydrogen column density, H$\alpha$ surface brightness, and star-formation rate based on resolved
stellar populations. We find that these tracers are indeed well described by powerlaws with a $-$1 index. This lends
credence to our model, and provides us with the crucial means to estimate the visibility time of SNRs in different
locations, which we review briefly in \S~\ref{sec:disc}, and more extensively in Paper~II, where we use it to derive the
SN rate and delay time distribution in the Clouds.  We conclude by summarizing our main results and outlining avenues
for future work in \S~\ref{sec:summary}.

\begin{figure*}
  \includegraphics[width=0.7\textwidth]{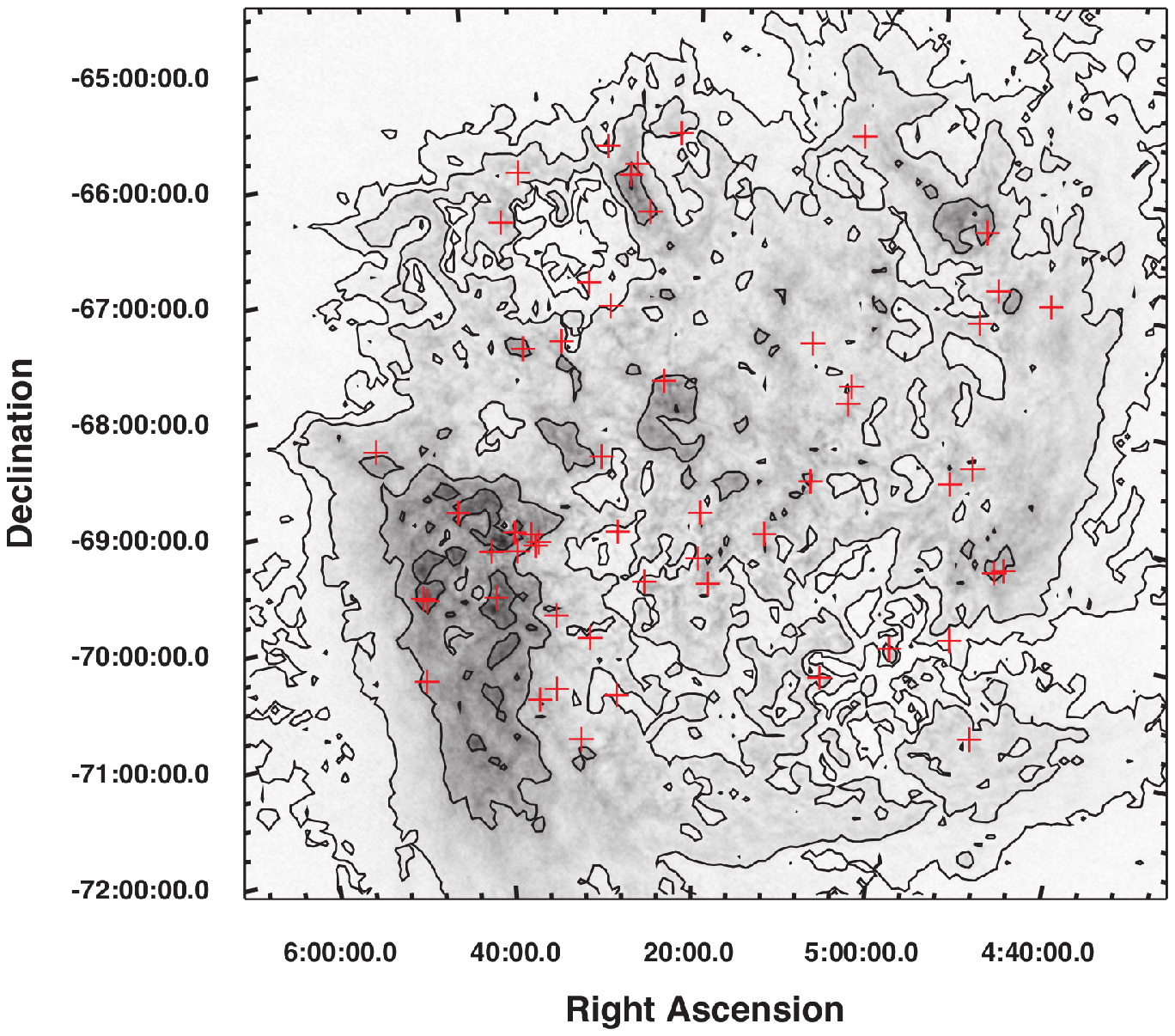}
  \includegraphics[width=0.7\textwidth]{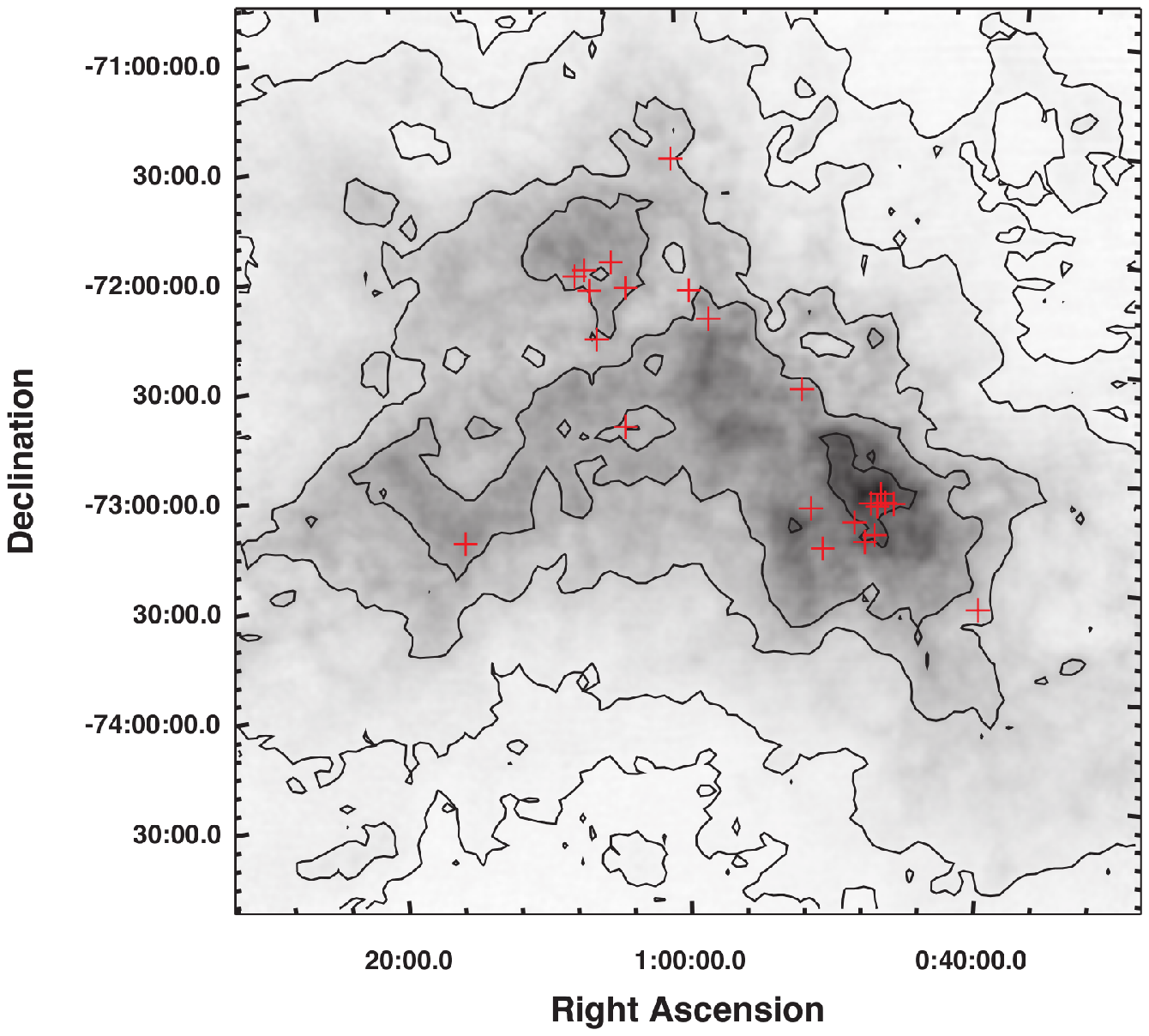}
  \caption{Maps of the LMC \citep[top, from][]{kim03:LMC_HI_Parkes_ATCA} and the SMC \citep[bottom,
    from][]{stanimirovic99MNRAS.302..417S} in the HI 21~cm line. The positions of the SNRs from Table 1
    are indicated by crosses: 54 objects in the LMC, 23 objects in the SMC. The contours are at column densities of
    5, 10, 30, 50, and $100\times10^{20}\,\mathrm{cm^{-2}}$. \label{fig-HIMaps}}
\end{figure*}

\section{The Supernova Remnants in the Magellanic Clouds: Observations and Sample Completeness}
\label{catalog}

The population of SNRs in the MCs has been the object of extensive study for many decades. Several catalogues have been
compiled at different wavelengths, from the radio to the optical and X-ray. Because SNRs in the Milky Way and the MCs
are usually discovered in the radio, it is the radio catalogues that often have the largest number of entries, but there
is some confusion in the literature regarding the total number of SNRs in the
Clouds. \citet{filipovic98:MC_Radio_Discrete} listed all the discrete radio sources in the Parkes survey of the MCs, and
found 32 SNRs and 12 SNR candidates in the LMC, and 12 SNRs in the SMC. According to \citet{payne08:LMCSNRs}, a revision
of the Parkes survey complemented with Australian Telecope Compact Array (ATCA) data yields 52 confirmed SNRs and 20
candidates in the LMC, but these sources are not listed in their paper. Instead, the authors present optical
spectroscopy of 25 of the 52 confirmed LMC SNRs. Data from ATCA was also collected for the SMC, where
\citet{filipovic05:SMc_SNRs} list 16 confirmed SNRs. An optical catalogue of MC SNRs is being assembled as part of the
Magellanic Clouds Emission Line Survey \citep[MCELS,][]{smith00:MCELS}; the most recent on-line version of this
catalogue\footnote{Last modified January 9 2006, see \url{http://www.ctio.noao.edu/mcels/snrs/framesnrs.html}} lists 31
SNRs in the LMC and 11 in the SMC. In the X-rays, \citet{williams99:LMC_SNR_Atlas} published an atlas of \textit{ROSAT}
sources, which contained 31 LMC SNRs. The most recent on-line version of this catalogue\footnote{Last modified August 29
  2006, see \url{http://www.astro.illinois.edu/projects/atlas/lmc_snr_pgs/lmctable.html}}, complemented with optical,
radio and infrared data, lists 38 confirmed SNRs in the LMC. Finally, \citet{heyden03:SMC_SNRs} present \xmm\
observations of 13 SMC SNRs.

Taken separately, these catalogues are not adequate for our purposes, because they disagree on the classification of
particular sources, and assign different names, sizes and positions to well-known objects. Catalogues at different
wavelengths can also be sensitive to SNRs at different stages of their evolution, so a multi-wavelength compilation is
the best way to obtain a global picture of the SNR population. At the writing of this paper, there is still no unified,
multi-wavelength catalogue of all the SNRs in the MCs in the refereed literature, although the Magellanic Clouds
Supernova Remnants (MCSNR) collaboration is assembling one \citep{2010AAS...21545416M} \footnote{For the most recent
  on-line version (2009), see \url{http://hoth.ccssc.org/mcsnr/}}. The current version of the MCSNR catalogue contains
48 confirmed objects in the LMC and 19 in the SMC. We have merged this on-line catalogue with the ones listed in the
previous paragraph, taking only the sources that are classified as confirmed SNRs in at least one catalogue, and
eliminating all duplicates and sources listed only as candidate SNRs. This yields 54 confirmed SNRs in the LMC and 23 in
the SMC, which we list in Table 1.  In Figure~\ref{fig-HIMaps}, we plot the positions of these 77 SNRs on the HI maps of
the MCs published by \citet{kim03:LMC_HI_Parkes_ATCA} and \citet{stanimirovic99MNRAS.302..417S}.

Because we are interested mostly in reliable sizes (to derive the histograms in \S~\ref{distribution}) and positions (to
calculate the delay time distribution in Paper~II), we take the centres and diameters of the SNRs from high resolution
X-ray observations (\ch\ or \xmm), whenever possible. If no high-resolution X-ray data are available, we revert to the
sizes and centers listed in the original catalogues, which were determined on a case-by-case basis, usually selecting
the wavelength that provided the cleanest measurement (see Table 1 and references for details). We have found that
high-resolution X-ray data provide the most reliable measurements for several X-ray bright and radio faint objects,
which are often small and have boundaries that are hard to resolve clearly using single-dish radio data. For the bulk
population, sizes determined at different wavelengths might disagree in individual cases, but this does not have a large
impact on the final distribution of sizes, as long as the data sets are of reasonable quality \citep[e.g., see Figure 7
and related discussion in][]{filipovic05:SMc_SNRs}.

\begin{table*}
 \centering
 \begin{minipage}{165mm}
  \caption{SNRs in the Magellanic Clouds}
  \begin{tabular*}{160mm}{@{\extracolsep{\fill}}llccccccc}
    \hline
    SNR\footnote{SNR names are constructed from the location of the centre in J2000 coordinates as given in the catalogues.}  &  Alternative  & \multicolumn{2}{c}{Position (J2000)}
    & Diameter & Wave- & Refe- & Radio flux & Catalogue \\
    & Name\footnote{Alternative names are given for clarification purposes; see \citet{williams99:LMC_SNR_Atlas}
      and \citet{filipovic05:SMc_SNRs} for a discussion of past conventions in SNR nomenclature.} & R.A. & Decl. &
    (arcsec) \footnote{Some catalogues give two axial diameters for asymmetric SNRs. In those cases, we list the mean
      diameter.} & length\footnote{For the center location and diameter. R: radio; O: optical; X: X-ray} &
    rence\footnote{For the center location and diameter. See table notes for a list of all the catalogues and other
      references.}  & density (Jy)\footnote{All flux densities are at 1.4 GHz from catalogue M, unless indicated.} & entries\footnote{Catalogues
      that list the object as a confirmed SNR; see table notes.} \\
    \hline
    \multicolumn{8}{c}{LMC SNRs}\\
    \hline
    J0448.4$-$6660 & \nodata & 04h 48m 22s & $-$66d 59m 52s & 220 & R & M & 0.04 & M \\	
    J0449.3$-$6920 & \nodata & 04h 49m 20s & $-$69d 20m 20s & 133 & R & M & 0.11 &  M \\
    J0450.2$-$6922 & B0450$-$6927 & 04h 50m 15s & $-$69d 22m 12s & 210 & R & P & 0.27\footnote{At 4.75 GHz, from F95} & P \\
    J0450.4$-$7050 & B0450$-$709 & 04h 50m 27s & $-$70d 50m 15s & 357 & X & W & 0.56 & WSFM \\
    J0453.2$-$6655 & N4 & 04h 53m 14s & $-$66d 55m 13s & 252 & X & W & 0.11 & WPFM \\      
    J0453.6$-$6829 & B0453$-$685 & 04h 53m 38s & $-$68d 29m 27s & 120 & X & G03 & 0.19 & WFSM \\
    J0453.9$-$7000 & B0454$-$7005 & 04h 53m 52s & $-$70d 00m 13s & 420 & R & P & 0.26\footnote{At 2.45 GHz, from F95} & P \\
    J0454.6$-$6713 & N9 & 04h 54m 33s & $-$67d 13m 13s & 177 & X & S06 & 0.06 & WFM \\
    J0454.8$-$6626 & N11L & 04h 54m 49s & $-$66d 25m 32s & 87  & X & W & 0.11 & WPFSM \\
    J0455.6$-$6839 & N86 & 04h 55m 37s & $-$68d 38m 47s & 348 & X & W & 0.26 & WPFSM \\
    J0459.9$-$7008 & N186D & 04h 59m 55s & $-$70d 07m 52s & 150 & O & W & 0.07 & WPFSM \\
    J0505.7$-$6753 & DEM L71 & 05h 05m 42s & $-$67d 52m 39s & 72 & X & B07 & 0.01 & WSM \\
    J0505.9$-$6802 & N23 & 05h 05m 55s & $-$68d 01m 47s & 111 & X & H06 & 0.35 & WPFSM \\
    J0506.1$-$6541 & \nodata & 05h 06m 05s & $-$65d 41m 08s & 408 & R & M & 0.11 & M \\
    J0506.8$-$7026 & B0507$-$7029 & 05h 06m 50s & $-$70d 25m 53s & 330 & R & P & 0.36\footnote{From F95} & P \\  
    J0509.0$-$6844 & N103B\footnote{This widely used name is the result of an incorrect identification, see entry in M} & 05h 08m 59s & $-$68d 43m 35s & 28 & X & B07 & 0.51 & WPFSM \\
    J0509.5$-$6731 & B0509$-$67.5 & 05h 09m 31s & $-$67d 31m 17s & 29 & X & B07 & 0.08 & WSM \\
    J0513.2$-$6912 & DEM L109  & 05h 13m 14s & $-$69d 12m 20s & 215 & R & Boj07 & 0.20 & WFSM \\
    J0518.7$-$6939 & N120 & 05h 18m 41s & $-$69d 39m 12s & 134 & X & R08 & 0.35 & WFM \\
    J0519.6$-$6902 & B0519$-$690 & 05h 19m 35s & $-$69d 02m 09s & 31 & X & B07 & 0.10 & WFSM \\
    J0519.7$-$6926 & B0520$-$694 & 05h 19m 44s & $-$69d 26m 08s & 174 & O & W & 0.12 & WPFSM \\
    J0521.6$-$6543 & \nodata & 05h 21m 39s & $-$65d 43m 07s & \nodata\footnote{No size is given for this SNR in the MCSNR catalogue.}
    & R & M & 0.03 & M \\
    J0523.1$-$6753 & N44 & 05h 23m 07s & $-$67d 53m 12s & 228 & R & K98 & 0.14 & WPFM \\
    J0524.3$-$6624 & DEM L175a & 05h 24m 20s & $-$66d 24m 23s & 234 & O & W & 0.11 & WPFSM \\
    J0525.1$-$6938 & N132D & 05h 25m 04s & $-$69d 38m 24s & 114 & O & W & 3.71 & WPFSM \\
    J0525.4$-$6559 & N49B & 05h 25m 25s & $-$65d 59m 19s & 168 & X & P03b & 0.32 & WPFSM \\	
    J0526.0$-$6605 & N49 & 05h 26m 00s & $-$66d 04m 57s & 84 & X & P03 & 1.19 & WPFSM \\
    J0527.6$-$6912 & B0528$-$692 & 05h 27m 39s & $-$69d 12m 04s & 147 & O & W & 0.05 & WPFSM\\
    J0527.9$-$6550 & DEM L204 & 05h 27m 54s & $-$65d 49m 38s & 303 & O & W & 0.14 & WPFSM \\
    J0527.9$-$6714 & B0528$-$6716 & 05h 27m 56s & $-$67d 13m 40s & 196 & R & M & 0.10 & FSM \\
    J0528.1$-$7038 & B0528$-$7038 & 05h 28m 03s & $-$70d 37m 40s & 60 & R & P & 0.28\footnote{At 2.45 GHz, from F95} & P \\
    J0529.1$-$6833 & DEM L203 & 05h 29m 05s & $-$68d 32m 30s & 667 & R & M & 0.24 & M \\
    J0529.9$-$6701 & DEM L214 & 05h 29m 51s & $-$67d 01m 05s & 100 & R & P & 0.06 & PM \\
    J0530.7$-$7008 & DEM L218 & 05h 30m 40s & $-$70d 07m 30s & 213 & R & M & 0.06 & M \\
    J0531.9$-$7100 & N206 & 05h 31m 56s & $-$71d 00m 19s & 192 & O & W & 0.33 & WFSM \\
    J0532.5$-$6732 & B0532$-$675 & 05h 32m 30s & $-$67d 31m 33s & 252 & X & W & 0.16 & WSM \\			                                                    
    J0534.0$-$6955 & B0534$-$699 & 05h 34m 02s & $-$69d 55m 03s & 114 & X & H03 & 0.08 & WPFSM \\
    J0534.3$-$7033 & DEM L238 & 05h 34m 18s & $-$70d 33m 26s & 180 & X & B06 & 0.06 & WPFSM \\
    J0535.5$-$6916 & SNR1987A & 05h 35m 28s & $-$69d 16m 11s & 2 & R & N08 & 0.05 & WM \\
    J0535.7$-$6602 & N63A & 05h 35m 44s & $-$66d 02m 14s & 66 & X & W03 & 1.43 & WPFSM \\
    J0535.8$-$6918 & Honeycomb & 05h 35m 46s & $-$69d 18m 02s & 102 & X & W & 0.30 & WFSM \\
    J0536.1$-$6735 & DEM L241 & 05h 36m 03s & $-$67d 34m 36s & 135 & X & Ba06 & 0.29 & WPFSM \\
    J0536.1$-$7039 & DEM L249 & 05h 36m 07s & $-$70d 38m 37s & 180 & X & B06 & 0.05 & WPFSM \\
    J0536.2$-$6912 & B0536$-$6914 & 05h 36m 09s & $-$69d 11m 53s & 480 & R & P & 4.26\footnote{From F95} & PF \\ 
    \hline
   \end{tabular*}
 \end{minipage}
\end{table*}

\begin{table*}
  \centering
  \begin{minipage}{165mm}
    \contcaption{}
    \begin{tabular*}{160mm}{@{\extracolsep{\fill}}llccccccc}
      \hline
      \multicolumn{8}{c}{LMC SNRs}\\
      \hline
      J0537.4$-$6628 & DEM L256 & 05h 37m 27s & $-$66d 27m 50s & 204 & R & M & 0.06 & M \\
      J0537.6$-$6920 & B0538$-$6922 & 05h 37m 37s & $-$69d 20m 23s & 169 & O & S & 1.01\footnote{At 4.75 GHz, from F95} & FS \\
      J0537.8$-$6910 & N157B & 05h 37m 46s & $-$69d 10m 28s & 102 & X & C06 & 2.64 & WFSM \\
      J0540.0$-$6944 & N159 & 05h 39m 59s & $-$69d 44m 02s & 78 & X & W00 & 1.90 & WM \\
      J0540.2$-$6920 & B0540$-$693 & 05h 40m 11s & $-$69d 19m 55s & 60 & X & H01 & 0.88 & WFSM \\
      J0543.1$-$6858 & DEM L299 & 05h 43m 08s & $-$68d 58m 18s & 318 & X & W & 0.21 & WSM \\
      J0547.0$-$6943 & DEM L316B & 05h 46m 59s & $-$69d 42m 50s & 84 & X & W05 & 0.52 & WFSM \\
      J0547.4$-$6941 & DEM L316A & 05h 47m 22s & $-$69d 41m 26s & 56 & X & W05 & 0.33 & WFM \\
      J0547.8$-$7025 & B0548$-$704 & 05h 47m 49s & $-$70d 24m 54s & 102 & X & H03 & 0.05 & WFSM \\
      J0550.5$-$6823 & \nodata  & 05h 50m 30s & $-$68d 22m 40s & 312 & R & M & 0.64 & M \\
      \hline
      \multicolumn{8}{c}{SMC SNRs}\\
      \hline  	
      J0040.9$-$7337 & DEM S5 & 00h 40m 55s & $-$73d 36m 55s & 121 & R & M & 0.19 & M \\
      J0046.6$-$7309 & DEM S32 & 00h 46m 39s & $-$73d 08m 39s & 136 & X & H &  0.07 & HMF05 \\
      J0047.2$-$7308 & IKT2 & 00h 47m 12s & $-$73d 08m 26s & 66 & X & H & 0.46 & HMF05 \\
      J0047.5$-$7306 & B0045$-$733 & 00h 47m 29s & $-$73d 06m 01s & 180 & R & F05 & 0.14\footnote{From F02} & F05 \\		
      J0047.7$-$7310 & HFPK419 & 00h 47m 41s & $-$73d 09m 30s & 90 & X & H & 0.14 & HM \\
      J0047.8$-$7317 & NS21 & 00h 47m 48s & $-$73d 17m 27s & 76 & R & F05 & 0.03 & F05 \\
      J0048.1$-$7309 & NS19 & 00h 48m 06s & $-$73d 08m 43s & 79 & R & F05 & 0.08 & F05 \\
      J0048.4$-$7319 & IKT4 & 00h 48m 25s & $-$73d 19m 24s & 84 & X & H & 0.15 & HMF05 \\
      J0049.1$-$7314 & IKT5 & 00h 49m 07s & $-$73d 14m 05s & 116 & X & H & 0.28 & HMF05 \\
      J0051.1$-$7321 & IKT6 & 00h 51m 07s & $-$73d 21m 26s & 144 & X & H05 & 0.10 & HMF05 \\
      J0051.9$-$7310 & IKT7 & 00h 51m 54s & $-$73d 10m 24s & 97 & X & M & 0.00\footnote{This SNR has no radio or optical counterpart} & M \\
      J0052.6$-$7238 & B0050$-$728 & 00h 52m 33s & $-$72d 37m 35s & 323 & R & F05 & 0.22 & MF05 \\
      J0058.3$-$7218 & IKT16 & 00h 58m 16s & $-$72d 18m 05s & 200 & X & H & 0.07 & HMF05 \\
      J0059.4$-$7210 & IKT18 & 00h 59m 25s & $-$72d 10m 10s & 158 & X & H & 0.37 & HMF05 \\
      J0100.3$-$7134 & DEM S108 & 01h 00m 21s & $-$71h 33m 40s & 149 & R & F05 & 0.21 & MF05 \\
      J0103.2$-$7209 & IKT21 & 01h 03m 13s & $-$72d 08m 59s & 62 & X & H & 0.10 & HMF05 \\
      J0103.5$-$7247 & HFPK334 & 01h 03m 30s & $-$72d 47m 20s & 86 & R & M & 0.05 & M \\
      J0104.0$-$7202 & B0102$-$7219 & 01h 04m 02s & $-$72d 01m 48s & 44 & X & G00 & 0.28 & HMF05 \\
      J0105.1$-$7223 & IKT23 & 01h 05m 04s & $-$72d 22m 56s & 170 & X & P03c & 0.09 & HMF05 \\
      J0105.4$-$7209\footnote{This source might be two smaller SNRs, see H, F05} & DEM S128 & 01h 05m 23s & $-$72d 09m 26s & 124 & X & H & 0.05 & HMF05 \\
      J0105.6$-$7204 & DEM S130 & 01h 05m 39s & $-$72d 03m 41s & 79 & R & F05 & 0.05 & F05 \\
      J0106.2$-$7205 & IKT25 & 01h 06m 14s & $-$72d 05m 18s & 110 & X & H & 0.01 & HMF05 \\
      J0114.0$-$7317 & N83C & 01h 14m 00s & $-$73d 17m 08s & 17 & R & M & 0.02\footnote{At 50 GHz} & M \\
      \hline 
    \end{tabular*}
    
    \textit{Catalogues:}
    
    \textbf{F}: \citet{filipovic98:MC_Radio_Discrete}. Catalogue of
    discrete radio sources in the MCs. Data from Parkes telescope.
    
    \textbf{F95}: \citet{filipovic95:LMC_Radio_5Bands}. Catalogue of discrete radio sources in the LMC, with fluxes at 1.40,
    2.45, 4.75, 4.85 and 8.55 GHz. Data from Parkes telescope.
    
    \textbf{F02}: \citet{filipovic02:SMC_radio_4bands}. Catalogue of discrete radio sources in the SMC, with fluxes at 1.42, 2.37,
    4.80, and 8.64 GHz. Data from ATCA and Parkes telescope. 

    \textbf{F05}: \citet{filipovic05:SMc_SNRs}. Catalogue of radio SNRs in the SMC. Data from ATCA. 

    \textbf{H}: \citet{heyden03:SMC_SNRs}. Catalogue of X-ray bright SNRs in the SMC. Data from \xmm.

    \textbf{M}: MCSNR on-line database, (R. Williams et al.) \url{http://hoth.ccssc.org/mcsnr/}.

    \textbf{P}: \citet{payne08:LMCSNRs}. Optical spectroscopy of radio SNRs in the LMC. Radio data from ATCA, optical
    data from Siding Spring Observatory and South African Astronomical Observatory.

    \textbf{S}: MCELS on-line catalogue (C. Smith et al.), \url{http://www.ctio.noao.edu/mcels/snrs/snrcat.html}.  Optical
    data from CTIO.

    \textbf{W}: \citet{williams99:LMC_SNR_Atlas}. X-ray atlas from \textit{ROSAT}, also including radio and optical
    observations. Available on-line at \url{http://www.astro.illinois.edu/projects/atlas/index.html}.
  
  \textit{Other References:}
  
  B06: \citet{borkowski06:DEML238_DEML249}; Ba06: \citet{bamba06:DEM_L241}; B07: \citet{badenes07:outflows}; Boj07:
  \citet{bojicic07:B0513}; C06: \citet{chen06:N157B}; G00: \citet{gaetz00:E0102};G03: \citet{gaensler03:0453}; H01:
  \citet{hwang01:N158A}; H03: \citet{hendrick03:LMC_SNRs}; H05: \citet{hendrick05:B0049}; H06: \citet{hughes06:N23};
  K98: \citet{kim98:N44}; N08: \citet{ng08:SN1987A}; P03: \citet{park03:N49}; P03b: \citet{park03:N49B}; P03c:
  \citet{park03:E0103}; R08: \citet{reyes-iturbide08:N120}; S06: \citet{seward06:N9}; W00: \citet{williams00:0540}; W03:
  \citet{warren03:N63A}; W05: \citet{williams05:DEML316}.
  
\end{minipage}
\end{table*}

\begin{figure}
  \includegraphics[width=0.32\textwidth, angle=90]{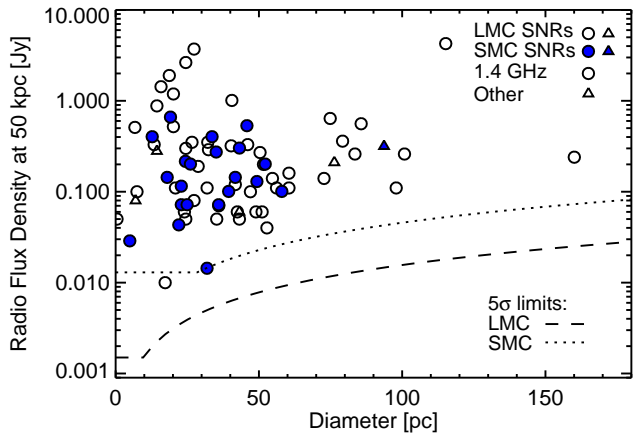}
  \caption{Integrated radio flux densities of SNRs in the LMC (empty symbols) and the SMC (filled symbols), plotted
    versus SNR diameter. The flux densities of the SMC SNRs have been multiplied by 1.44 for comparison. The circles
    indicate flux densities at 1.4 GHz, the triangles are at higher frequencies (see Table 1 for details).  The dashed
    and dotted lines show conservative $5\sigma$ limits for the selection of candidate SNRs in both galaxies, accounting
    for the flux limit per beam and the number of beams per object of given diameter
    \citep[see][]{hughes07:LMC_21cm_ATCA,filipovic02:SMC_radio_4bands}. The SMC limit has also been scaled by
    1.44. There is a real paucity of SNRs below a flux density of $\sim 50$~mJy at 50 kpc, corresponding to a luminosity
    of $\sim 1.5\times 10^{15}$~W~Hz$^{-1}$. \label{fig-radiofluxes}}
\end{figure}

With only one exception (SNR J0051.9$-$7310 in the SMC, see Table 1), all the SNRs in the MCs are detected in the radio,
and their flux densities can be found either in the MCSNR catalogue or in the general multiband radio source catalogues of
\citet{filipovic95:LMC_Radio_5Bands} and \citet{filipovic02:SMC_radio_4bands}. We have listed these flux densities in
Table 1, at 1.4 GHz when available, and at higher frequencies when there was no 1.4 GHz detection. In
Figure~\ref{fig-radiofluxes}, we plot the radio flux densities as a function of SNR diameter, adopting the fiducial
distances of 50 kpc to the LMC \citep{alves04:LMC_Distance} and 60 kpc to the SMC \citep{hilditch05:SMC_Distance} to
convert angular sizes to linear dimensions (we will use these distances for the remainder of the paper). The flux
densities of the SMC SNRs have been multiplied by a factor 1.44 for comparison to the LMC SNRs. Remarkably, there
appears to be a ``floor'' in SNR flux density at $\sim 50$~mJy, with only a few objects being fainter than this
limit.

The degree of completeness of our sample is hard to estimate, because we include objects that were found using a
very hetrogeneous mix of selection criteria, but a number of factors indicate that we should not be missing a large
number of SNRs. First, the floor at 50~mJy is much brighter than the nominal sensitivity of the ATCA/Parkes radio
surveys, which at 1.4 GHz is around 0.3~mJy per beam in the LMC \citep[][]{hughes07:LMC_21cm_ATCA}, and 1.8~mJy per beam
in the SMC \citep{filipovic02:SMC_radio_4bands}, with beam sizes of $40''$ and $98''$, respectively. Even requiring at
least a $5\sigma$ detection for the selection of SNR candidates, and allowing for the fact that large SNRs should be
harder to identify at the same level of significance, the vast majority of our objects are safely above any conservative detection
limits in the radio (see Figure~\ref{fig-radiofluxes}). Second, the systematic searches for SNRs on the ATCA/Parkes
surveys by \citet{filipovic05:SMc_SNRs} and \citet{payne08:LMCSNRs} have not uncovered a substantial population of
previously unknown SNRs between the radio floor and the detection limits. We note that the total number of confirmed
SNRs reported (but not listed) by \citet{payne08:LMCSNRs} in the LMC is very close to our own count (52 vs. 54). Third,
the existence of a substantial population of radio detected SNRs that are masquerading as other kinds of sources is
unlikely in the presence high quality multi-wavelength data. Small, radio-faint SNRs that might be mistaken for
background sources are often young objects with bright X-ray emission like SNRs J0509.5$-$6731 (B0509$-$67.5) or
J0505.7$-$6753 (DEM L71). Older, more extended objects that might be confused with surrounding or nearby HII regions are
usually identified at optical wavelengths by diagnostic quantities like the [S II]/H$\alpha$ ratio, which can be used to
distinguish shocked nebulae from photoionized gas \citep{fesen85:Evolved_SNRs}.

A related concern is that some SNRs might be ``lost'' inside superbubbles formed by several nested SN explosions
\citep{maclow88:superbubbles}. In practice, SNRs evolving inside low-density cavities do not disappear -- they just
become fainter and evolve more slowly, at least until the blast wave reaches the cavity edges. The largest, most
prominent superbubble complex in the LMC, 30 Doradus, harbors several well-studied SNRs, including J0537.8$-$6910
(N157B) \citep{townsley06:30Dor}, and many other objects listed in Table 1 are associated with superbubbles. As long as
the surveys are sensitive enough to find faint SNRs, this should not be a major issue.

In the absence of a systematic re-analysis of all the available data with a set of homogeneous criteria to identify SNRs
in both Clouds, we cannot guarantee that our sample is $100\%$ complete. Such an analysis is outside the scope of the
present work, but all the evidence seems to indicate that our compilation should not be missing a large number of
objects, and it should include few, if any, spurious ones. We conclude that our list of MC SNRs is at least \textit{fairly}
complete (which is enough for our goals), and that the paucity of SNRs below $\sim 50$~mJy is probably real, and not due
to any observational selection effects. This suggests that any MC SNRs with flux densities between the radio floor and
the detection limits must fade relatively quickly. As we shall discuss in \S~\ref{sec:disc}, the results of Paper II can
be used to argue independently that the SNR sample we have compiled has a high degree of completeness.

\section{The distribution of remnant sizes}
\label{distribution}

\begin{figure*}
\includegraphics[width=0.32\textwidth,angle=90]{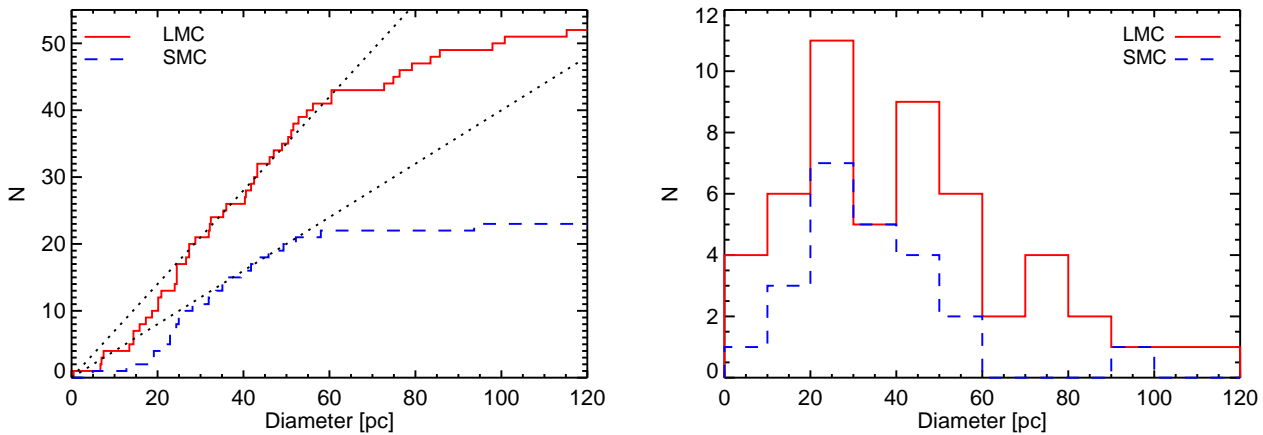}
\caption{Distribution of the SNR sizes listed in Table 1 for the LMC (red solid plots) and SMC (blue dashed plots),
  represented as cumulative (left) and differential (right) histograms. For illustrative purposes, linear size
  distributions normalized to give the correct number of SNRs at a diameter of 40 pc have been overplotted as
  dotted lines on the cumulative histograms. Both galaxies have SNR size distributions that are close to linear in the
  cumulative, or flat in the differential, up to a cutoff at $r\sim30$ pc. }
\label{sizedist}
\end{figure*}

Figure~\ref{sizedist} shows the cumulative and differential histograms of remnant sizes in the LMC and in the SMC. In
both Clouds, the cumulative distributions are roughly linear in remnant size, i.e., with an equal number of remnants per
equal size bin in the differential distributions, up to a cutoff at a physical radius $r_{\rm cut}\sim 30$~pc.

Because the data points in a cumulative histogram are not independent of one another, these curves cannot be fitted in
the usual way (using the $\chi^{2}$ statistic). A robust, quantitative estimate for the slope of the distributions can
be obtained by performing instead a maximum-likelihood fit, as follows \citep[see][for a similar treatment applied to a
different problem]{maoz93:grav_lensing_statistics}. Suppose that a particular model predicts a size distribution
$dN/dr=n(r)$, which integrates to $\int _0^{r_{\rm cut}} n(r) dr = N$, where $N$ is the total number of remnants up to
$r_{\rm cut}$.  If we bin our data into many small bins between $r=0$ and $r_{\rm cut}$, each of width $\delta r$, most
bins will have zero SNRs, and some will have one SNR. Given the model, the Poisson probability of finding $j$ remnants
in the $i$th bin, for which the model predicts $n(r_i)$ remnants, is
\begin{equation}
P(j|n(r_i))=e^{-n(r_i)}n(r_i)^{j}/j ! .
\end{equation}
We will define the likelihood of a given model as the product of these probabilities. The logarithm of the likelihood,
considering that $j$ always equals either 0 or 1, simplifies to
\begin{equation}
\ln L =[\sum_{i (j>0)} \ln n(r_i)] - N,
\label{loglike}
\end{equation}
where the summation is only over the specific data values of the SNR radii.  A power law of index $\alpha$, having the
proper normalization, will have the form
\begin{equation}
n(r)=\frac{N (\alpha+1)}{r_{\rm cut}^{\alpha+1}}r^\alpha.
\end{equation}
Inserting in Eq.~\ref{loglike}, differentiating $\ln L$ with respect to $\alpha$, and equating to zero to find the
maximum gives the maximum likelihood solution,
\begin{equation}
(\alpha+1)_{\rm ml}=\frac{N}{N\ln r_{\rm cut} - \sum_{j>0} \ln r_i},
\end{equation}
with an uncertainty on $\alpha$ of
\begin{equation}
\Delta \alpha=\left(-\frac{d^2(\ln
 L)}{d\alpha^2}\right)^{-1/2}=\frac{\alpha+1}{\sqrt N}.
\end{equation}

This procedure yields a maximum likelihood index of $\alpha=0.14\pm 0.18$ for the LMC, and $\alpha=0.32\pm 0.28$ for the
SMC. Thus, the LMC appears to indeed have a SNR size distribution that is close to uniform. In the SMC, the best fit is
intermediate between a flat distribution and one that rises linearly with radius, but given the smaller number of SNRs,
it is consistent with both slopes. From Figure~\ref{sizedist}, it appears that the steeper slope in the SMC is driven by
the small-radius side of the distribution. Indeed, if we fit separately the first 8 points and the following 14 points,
the maximum likelihood solution is $\alpha=1.7\pm 1.0$ at small radii, and $\alpha=0.17\pm 0.23$ thereafter.  This
result confirms the visual impression, but formally it is still consistent with a slope close to zero at all radii at
the $1.7\sigma$ level. We conclude that the SNR size distribution in both Clouds is consistent with being roughly
uniform, although there are hints for a deviation at small radii in the SMC.  It is possible that a deficit at small
radii is also present in the LMC distribution (see Figure \ref{sizedist}), but the Poisson errors are too large to claim
that the data require it.
 
The linear cumulative distribution of SNR sizes in the MCs (and also the Milky Way) has been previously noted and
discussed by \citet{mathewson84:SNR-Magellanic-Clouds}, \citet{mills84:Radio_SNRs}, \citet{green84:SNR_Statistics},
\citet{hughes84:SNR_N_D}, \citet{fusco-femiano84:LMC_SNRs}, \citet{berkhuijsen87:SNRs_N_D},
\citet{chu88:LMC_SNRs_environments}, and most recently by \citet{bandiera10:Statistics_SNRs}.  Several of these papers
also pointed out cutoffs in the distribution.  All of these papers considered smaller SNR samples, often based on much
shallower radio data and, with few exceptions, did not include multi-wavelength observations. Some of these authors
interpreted the observed size distribution as evidence that most MC SNRs are in their ``free expansion'' phase, during
which the shock velocity is constant, and inferred that these SNRs expand into an extremely low-density
medium. Alternatively, \citet{green84:SNR_Statistics} and \citet{hughes84:SNR_N_D} warned that the observed distribution
was the result of selection effects; most objects they discussed had been selected in X-rays, and the X-ray flux limits
then led to the exclusion of larger and fainter remnants, and their faint radio counterparts. Our present compilation is
bigger, it incorporates the most recent multi-wavelength data, it extends to larger sizes, and most importantly, it is
sensitive to radio flux densities two orders of magnitude below the observed luminosity floor. With these data we now
confirm the luminosity floor, the uniform size distribution, and the cutoff at $r_{\rm cut} \sim 30$~pc in the
Magellanic Cloud SNRs.

\begin{figure*}
\includegraphics[width=0.32\textwidth,angle=90]{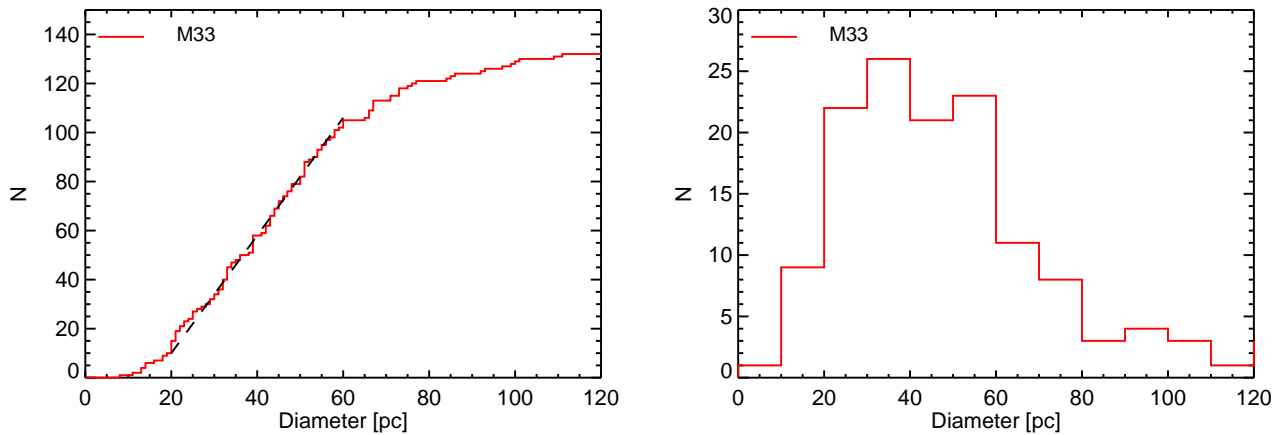}
\caption{Distribution of SNR sizes in M33 from \citet{long10:M33_SNRs}, represented as cumulative (left) and differential
  (right) histograms. A linear size distribution for diameters between 20 and 60 pc, normalized to give the correct
  number of SNRs at a diameter of 40 pc, has been overplotted as a dashed line on the cumulative histogram.}
\label{sizedistM33}
\end{figure*}

These features of the SNR size distribution are also present in other galaxies. Due to its proximity and face-on
orientation, M33 has probably the best SNR sample outside of the MCs. The most recent catalogue of M33 SNRs has been
published by \cite{long10:M33_SNRs}, and it includes data in the radio, optical, and X-rays \citep[from the ChASeM33
survey by \textit{Chandra},][]{plucinsky08:ChASeM33}. The distribution of SNR sizes given in Table 3 of
\cite{long10:M33_SNRs}, which we plot in Figure \ref{sizedistM33}, has the same broad characteristics we find in the
Clouds, but backed by much better statistics (137 objects). The cutoff at a physical radius of $r\sim30$ pc is very
clear, and the distribution appears remarkably uniform for radii between $\sim10$ pc and the cutoff. In this size range,
our method to estimate the maximum likelihood index yields $\alpha=0.04\pm 0.11$. Graphically, a linear function with
the correct normalization reproduces the cumulative histogram quite well (see Figure \ref{sizedistM33}). A significant
deficit of SNRs at radii below 10 pc is also obvious in the data. \cite{long10:M33_SNRs} warn that a few small (young)
SNRs might have escaped detection (see their Section 9), but given the numbers involved, most of the deficit may be
real. This indicates a deviation at small diameters from the uniform distribution along the lines of the one we noted in
the MCs.

The available data seem to indicate that the distribution of SNR sizes in M33 and the Magellanic Clouds is indeed close
to uniform between $r\sim10$ pc and a sharp cutoff at $r_{\rm cut}\sim30$ pc. In the case of M33, this conclusion needs
to be validated by a careful assessment of the completeness of the SNR sample, although from the discussion in \S~7.3 of
\cite{long10:M33_SNRs} it seems unlikely that a large number of M33 SNRs have escaped detection. Unfortunately, SNR
samples of this quality are hard to obtain for other galaxies, so it is difficult to verify how widespread these
features of the SNR size distribution are in reality. We will return briefly to this issue in \S~\ref{sec:summary}. In
the following section, we argue that a uniform size distribution, including a deficit of SNRs at small radii and a sharp
cutoff at a certain radius, arises naturally from the physics of SNR evolution and the distribution of densities in the
interstellar medium.
 
\section{Physics behind the size distribution}
\label{physics}

The evolution of SNRs is usually separated into three phases \citep[e.g.][]{woltjer72:SNR-review}. In the first,
free-expansion, phase, the mass swept up by the expanding shock is small compared to the ejected mass, and the shock
velocity is constant with time. In the Sedov-Taylor\footnote{The well-known solution for the propagation of a blast wave
  following a point-like explosion, commonly attributed to Leonid Sedov and Geoffrey Taylor, was also derived
  independently by John von Neumann.} phase, the swept-up mass is larger than the ejecta mass, and the expansion
decelerates. However, the cooling time of the shocked gas is still longer than the age of the remnant, and the evolution
is approximately adiabatic. Once the cooling time becomes comparable to the age, the SNR quickly loses energy
radiatively, entering the radiation-loss-dominated snowplough phase, after which it slows down, breaks up, and merges
with the interstellar medium.

Quantitative estimates show that typical SNRs should be in their Sedov-Taylor phases for ages between a few and a few
tens of kyrs, and for sizes of order a few to a few tens of parsecs
\citep[e.g.][]{cioffi88:Radiative_SNRs,blondin98:Radiative_SNRs,truelove99:adiabatic-SNRs}. Thus, most of the MC
SNRs in our sample should be in the Sedov-Taylor phase of their expansions, with radii growing as
\begin{equation}
\label{sedov}
r\sim E_0^{1/5}\rho^{-1/5}t^{2/5}, 
\end{equation}
where $E_0$ is the kinetic energy of the explosion, $\rho$ is the ambient gas density, and $t$ is the time.  The shock
velocity therefore decreases as
\begin{equation}
\label{sedovv}
v=\frac{dr}{dt}\sim E_0^{1/5} \rho^{-1/5}t^{-3/5},
\end{equation}
or equivalently expressed in terms of $r$ rather than $t$,
\begin{equation}
\label{vsimrhor}
v\sim E_0^{1/2} \rho^{-1/2}r^{-3/2}.
\end{equation}
Reasonably assuming a constant SN rate in the LMC over the past few kyrs, $dN/dt={\rm const.}$, the naively expected
size distribution of SNRs in their Sedov phase is
\begin{equation}
\frac{dN}{dr}=\frac{dN}{dt}\frac{dt}{dr}\sim r^{3/2},
\end{equation}
where we have used Eq.~\ref{vsimrhor} to substitute for $dt/dr$. This is contrary to the observed nearly uniform
($dN/dr\sim r^0$) distribution.  As noted above, this has been interpreted before either as evidence for free expansion,
or for selection effects, where fading of the older (and hence larger) remnants takes them below the detection limits,
and culls the pileup in the distribution expected at large sizes for a decelerating population. Both explanations are
unlikely for our MC SNR sample.
 
We propose, instead, that the uniform size distribution arises as a result of the transition from the Sedov phase to the
radiative phase \citep[see also][]{fusco-femiano84:LMC_SNRs,bandiera10:Statistics_SNRs}.  The radius of this transition
depends on ambient density, with SNRs evolving in dense environments becoming radiative sooner than SNRs in low-density
regions. When coupled with the distribution of densities in the Clouds, this leads to fewer and fewer sites at which
large-radius Sedov-phase SNRs can exist. We show this as follows.

The cooling time of the shocked gas depends on the density as
\begin{equation}
t_{\rm cool}\sim  \frac{kT}{\rho \Lambda(T)}
\end{equation}
where $\Lambda(T)$ is the cooling function at temperature $T$.  Following, e.g., \citet{mckee77:ISM_Theory},
\citet{blondin98:Radiative_SNRs}, or \citet{truelove99:adiabatic-SNRs}, $\Lambda(T)$ can be approximated with piecewise
power laws, with a dependence of $T^\epsilon$ in the temperature range of relevance for the shocked gas, around
$10^6~{\rm K}$. By relating the temperature to the shock velocity $v$, $kT\sim m_p v^2$ (where $m_p$ is the proton
mass), and equating $t_{\rm cool}$ to the age $t$ as expressed in Eq.~\ref{sedovv}, one obtains that the transition
radius, $r_{\max}$, scales as \citep[e.g.][]{bandiera10:Statistics_SNRs}
\begin{equation}
\label{rmaxwithepsilon}
r_{\rm max}\sim  E_0^{(3-2\epsilon)/(11-6\epsilon)}\rho^{-(5-2\epsilon)/(11-6\epsilon)}.
\end{equation}
A remnant expanding beyond this radius, at the given ambient density, will enter the radiative phase and quickly fade
from view.  For a fairly large range of plausible cooling function dependences, e.g., indices $\epsilon$ of $-1/2$ to
$-3/2$, $r_{\rm max}$ depends on density as $\rho^{-3/7}$ to $\rho^{-2/5}$. Conversely, the maximum ambient density that
will permit a Sedov-phase SNR of radius $r$ is
\begin{equation}
\rho_{\rm max}\propto  r^{\delta},
\label{rhomaxvsr}
\end{equation}
where $\delta$ is likely in the range $-7/3$ to $-5/2$.  
  
The expected size distribution of remnants will now be
\begin{equation}
\label{dNdr}
\frac{dN}{dr}=\frac{dN}{dt}\frac{dt}{dr}\sim r^{3/2}\int_{\rho_{\rm
    min}}^{\rho_{\rm
    max}} \rho^{1/2}\frac{dP}{d\rho}d\rho, 
\end{equation}
where $dP/d\rho$ is the distribution of gas densities in a given galaxy, and ${\rho_{\rm min}}$ is the minimum density
in that galaxy, on the size scales relevant for SNRs.  Let us parametrize the density distribution as a power law of
index $\beta$,
\begin{equation}
\frac{dP}{d\rho}\sim \rho^\beta , 
\end{equation}
with $\beta > -3/2$.  Then, substituting in Eq.~\ref{dNdr}, integrating, and expressing $\rho_{\rm max}$ in terms of $r$
using Eq.~\ref{rhomaxvsr}, gives a remnant size distribution
\begin{equation}
\frac{dN}{dr}\sim r^{\delta(\beta+3/2)+3/2}.
\end{equation}
A uniform size distribution, $dN/dr\approx {\rm const.}$, is obtained if $\beta=-6/7$ (for $\delta=-7/3$) or
$\beta=-9/10$ (for $\delta=-5/2$). In other words, the observed, roughly uniform, SNR size distribution implies a
powerlaw distribution of densities with index $\beta\approx -1$.

We note that, because of the dependence of $r_{\rm max}$ on $t_{\rm cool}$, $r_{\rm max}$ can depend on the metallicity
of the gas, $Z$. Assuming a cooling function proportional to $Z$, Eq. \ref{rmaxwithepsilon}, in addition to the
dependence on $\rho$ and $E_0$, will depend on $Z$ as
\begin{equation}
r_{\rm max}\sim  Z^{-2/(11-6\epsilon)}.
\end{equation}
However, in dwarf galaxies like the MCs, where $Z$ does not change much from one location to another, this will be a
small effect. For example, if $\epsilon=-1$, changing $Z$ by a factor 5 would only change $r_{\rm max}$ by $\sim20\%$.

The existence of a cutoff in the distribution beyond some size is naturally expected in the above picture, given the
existence of some minimum density $\rho_{\rm min}$ in the regions of the galaxies where SNe explode, and in view of the
decrease of $\rho_{\rm max}$ with radius. At some radius, $\rho_{\rm max}$ will equal $\rho_{\rm min}$ and the integral
in Eq.~\ref{dNdr} will therefore become zero. In other words, there will be nowhere in the MCs a region with a density
low enough to permit a SNR of that size that is still in its bright Sedov phase.

A deficit of SNRs at small radii, as observed in M33 and the MCs, is also expected in this scenario. Before the
onset of the Sedov stage, faster shock velocities will lead to fewer objects observed in the bins with the smallest
radii. This onset happens at ages (sizes) that depend on the details of the ejecta structure, as well as the ambient
density \citep[see \S~7 in][]{truelove99:adiabatic-SNRs}, but for most SNRs it should occur around a few hundred years
(a few pc), which is consistent with the deficits that we have observed in M33 and the MCs.  This regime affects only a
small number of objects in the MCs, and does not impact any of the arguments made above, so we will ignore it for the
remainder of the paper.

We have also ignored the deviations from the standard evolutionary picture that can be introduced by the shape of the
circumstellar medium excavated by the SN progenitors. \citet{badenes07:outflows} showed that most Type Ia SNRs with
known ages have sizes that are consistent with an interaction with a uniform ambient medium, but no such study has been
done for CC SNRs. The fast stellar outflows expected from the more massive CC SN progenitors will modify the sizes of a
few individual SNRs at certain stages of their evolution
\citep[e.g.,][]{dwarkadas05:SNR-Bubbles_1D,dwarkadas07:SNRs_Bubbles_WR}, but most of the objects that we consider here
are too large to be expanding in even the most extreme wind-blown cavities. As long as the bulk of the SNRs in the
sample spend most of their lifetimes in the Sedov stage, this should not affect our scenario. Similarly, the fact that
some SNRs evolve inside superbubbles \citep[e.g.][]{maclow88:superbubbles} is naturally incorporated into our picture --
superbubbles merely become one more of the factors driving the density distribution in the interstellar
medium. Incidentally, the size distribution of superbubbles also relates to the properties of the interstellar medium,
as shown by \citet{oey97:superbubble_sizes} for several nearby galaxies, including the SMC.
 
\section{Three estimates of the gas density distribution in the Magellanic Clouds}
\label{densityestimates}

As we have seen, a uniform SNR size distribution can be understood as the result of Sedov expansion, combined with a
transition to the radiative phase at an age that depends on the local density, provided that the density of the gas in
the interstellar medium follows a distribution close to a power law with an index of $-1$, $dP/d\rho\sim \rho^{-1}$. In
this Section, we test this hypothesis by examining three indirect tracers of gas density in the Magellanic Clouds: HI
column density; star-formation rate (SFR) based on resolved stellar populations; and H$\alpha$ emission-line surface
brightness. These tracers are well suited for our goals because they are valid over a wide range of densities, and the
necessary data are available from public surveys that cover the whole extent of the Clouds, as described in detail
below.


\subsection{21~cm emission-line-based HI column density}
\label{sec:HI}

\begin{figure}
\includegraphics[width=0.32\textwidth, angle=90]{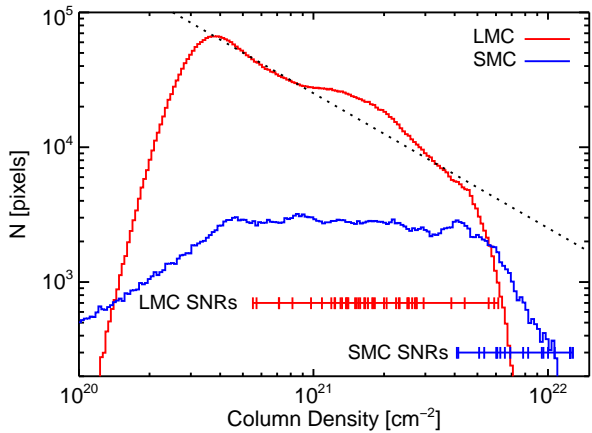}
\caption{Binned histogram of neutral H column densities in the LMC (red) and the SMC (blue). The dotted line overlaid on
  the LMC histogram is a power law of index $-1$ that intersect the data at a column of $5\times10^{20}\,\rm{cm^{-2}}$. The column
  density values for the cells that contain the SNRs in both galaxies have been represented by tick marks on the
  horizontal rulers.}
\label{HIHist}
\end{figure}

We have taken the surface brightness of HI 21~cm line emission in the MCs from the maps of
\citet{kim03:LMC_HI_Parkes_ATCA} and \citet{stanimirovic99MNRAS.302..417S}, which combine single-dish Parkes and
aperture-synthesis ATCA data to probe both small and large scales in the LMC and the SMC, respectively. The 21~cm
emission is optically thin, so the surface brightness is directly proportional to the HI column density. Since the LMC
possesses a fairly face-on \citep[inclination $i\sim 35^{\circ}$,][]{vandermarel01:LMC_inclination}, well-ordered HI
disk, the column density should, in turn, be roughly proportional to the volume density
$\rho$. \citet{kim07:HI_Clouds_LMC} report that the HI column density distribution of individual ``clouds'' of neutral
hydrogen in the LMC follows a log-normal form, rather than a power law. However, their figure~13 suggests that, above a
low cutoff of $2\times 10^{20}~{\rm cm}^{-2}$, the distribution does behave as a power law of slope $\sim -1$, over at
least an order of magnitude. To re-examine this, in Figure \ref{HIHist} we show the differential distribution of HI
column density in the individual beam-sized pixels of the \citet{kim03:LMC_HI_Parkes_ATCA} LMC map \citep[as opposed to
the cumulative plot for ``clouds'' shown in][]{kim07:HI_Clouds_LMC}.  We see that the HI column in the LMC does follow
an index $-1$ power law fairly well, between a column of $3\times 10^{20}~{\rm cm}^{-2}$ and $6\times 10^{21}~{\rm
  cm}^{-2}$.  The observed low cutoff in the column density is unavoidable because of the integration through the disk
and over the beam size (every line of sight is basically sampling the densest regions at that point). In regions with
low density, the H atoms might be ionized, as in the ``warm ionized'' phase of the interstellar medium
\citep{ferriere01:ISM}, so the tracer may become less reliable there. It is quite plausible that, in the regions where
SNe actually explode, the underlying distribution of densities also reaches a minimum -- as we recall, a minimum density
is required in order to reproduce the observed upper cutoff in SNR size -- but this does not necessarily correspond with
the lower threshold in the HI distribution. To illustrate this, we have calculated the mean HI columns in each of the
spatial ``cells'' defined by \citet{harris04:SMC_SFH} and \citet{harris09:LMC_SFH} that contain SNRs (see
\S~\ref{sec:SFR} below for a description of the cells), which we display with the horizontal rulers in Figure
\ref{HIHist}. We note that, in the LMC, these cells have average column values between $5\times10^{20}~{\rm cm}^{-2}$
(close to, but higher than the low HI cutoff) and $6\times 10^{21}~{\rm cm}^{-2}$, although most SNRs appear clustered
around $2\times 10^{21}~{\rm cm}^{-2}$.

As shown in Figure \ref{HIHist}, the distribution of HI column densities in the SMC is flat over the same range,
although the rise and fall from the plateau happen at the same densities as the rise and fall of the powerlaw in the
LMC. While we do not know the reason for this, we speculate that it may be related to an SMC geometry that is elongated
along our line of sight, and the integration effect that results.  The actual ``depth'' of the SMC, whether just a few
kpc or as much as 20~kpc, is debated
\citep{hatzidimitriou89:SMC_structure,harris04:SMC_SFH,subramanian09:LMC_SMC_Depth}, but is likely at least a few times
larger than that of the nearly face-on LMC. Such an integration effect would explain why the SMC SNRs are found at HI
column densities that are a factor $\sim$4 higher than the LMC SNRs (see horizontal rulers in Figure \ref{HIHist}). Two
other tracers of density, examined below, do suggest a $-1$ power law in the SMC, as well as in the LMC, and do not
display this offset in SNR ambient densities.


\subsection{Schmidt Law plus star formation rates from resolved stellar populations}
\label{sec:SFR}

As a second way to estimate the local gas densities and test the powerlaw density distribution hypothesis, we use star
formation rates (SFRs) calculated from the star-formation history (SFH) maps of the MCs published by
\citet{harris04:SMC_SFH} and \citet{harris09:LMC_SFH}, which are based on resolved stellar populations\footnote{The
  complete maps are available at \url{http://ngala.as.arizona.edu/dennis/mcsurvey.html}.}. The maps were elaborated
using four-band (\textit{U, B, V}, and \textit{I}) photometry from the Magellanic Clouds Photometric Survey
\citep{zaritsky04:MCPS}, which has a limiting magnitude between 20 and 21 in \textit{V}, depending on the local degree
of crowding in the images. In each Cloud, the data were divided into regions or ``cells'' with enough stars to produce
color-magnitude diagrams of sufficient quality, which were then fed into the StarFISH code \citep{harris01:StarFISH} to
derive the local SFH for each cell. For the LMC, \citet{harris09:LMC_SFH} divided more than 20 million stars into
spatial cells encompassing the central 8$^\circ\times8^\circ$ of the galaxy (see their figure 4). In total, they
produced 1376 cells for the LMC, most of them $12'\times12'$ in size, and about 50 cells in regions of lower stellar
density with a larger size ($24'\times24'$).  For the SMC, \citet{harris04:SMC_SFH} divided over 6 million stars into
351 $12'\times12'$ cells, leaving out two areas that are contaminated by Galactic globular clusters in the foreground
(see their figure 3). 

\citet{harris04:SMC_SFH} and \citet{harris09:LMC_SFH} provide the SFH for each of these cells, which we use in
Paper~II to derive the delay time distribution of SNe in the Clouds, but for our present purposes we are only interested
in the current SFR, which we have obtained by averaging the star formation in each cell over the last 35~Myr. The
\citet{schmidt59:SFR} law relates star formation to gas mass column, $\Sigma$, roughly as ${\rm SFR}\propto \Sigma^{3/2}$.
If gas density is distributed as $\rho^\beta$, and $\Sigma \propto \rho$, then SFR will be distributed as a power law of
index $2(\beta -1/2)/3$. Coincidentally, if $\beta=-1$, the SFR distribution will also have an index $-1$ powerlaw
distribution.  \citet{kennicutt98:SF_review} gives an update of the Schmidt law, relating star formation rate
surface density, $\Sigma_{\rm SFR}$, to gas mass column $\Sigma_{\rm gas}$, as
\begin{equation}
\Sigma_{\rm SFR}=(2.5\pm0.7)\times 10^{-4}\left(\frac{\Sigma_{\rm
    gas}}{M_\odot~{\rm pc}^{-2}}\right)^{1.4\pm0.15}M_\odot ~{\rm yr}^{-1}
{\rm kpc}^{-2}.
\label{schmidtlaw}
\end{equation}
As emphasized by \cite{kennicutt89:SFR_galdisks}, the Schmidt law has a threshold at a given mass column, below which
the star-formation rate falls steeply. The exact value of this threshold is disputed: \cite{kennicutt89:SFR_galdisks}
finds $\Sigma_{\rm gas}=3-4~{\rm M_{\odot}\,pc}^{-2}$, which corresponds to $\sim1.4\times10^{-3}~{\rm
  M_{\odot}\,yr^{-1}\,kpc}^{-2}$, but higher and lower threshold values have also been reported in the literature
\citep[see][for an updated discussion of this issue]{bigiel08:SF_law_sub_kpc}. From Eq.~\ref{schmidtlaw} with
$\Sigma_{\rm gas}=3.5~M_\odot {\rm pc}^{-2}$, the threshold mass column corresponds to a SFR of $0.0014 ~M_\odot~ {\rm
  yr}^{-1} {\rm kpc}^{-2}$, or $4.3\times10^{-5}~ M_\odot~ {\rm yr}^{-1}$ per $12'\times 12'$ cell in the
\citet{harris09:LMC_SFH} maps of the LMC.

\begin{figure}
\includegraphics[width=0.32\textwidth, angle=90]{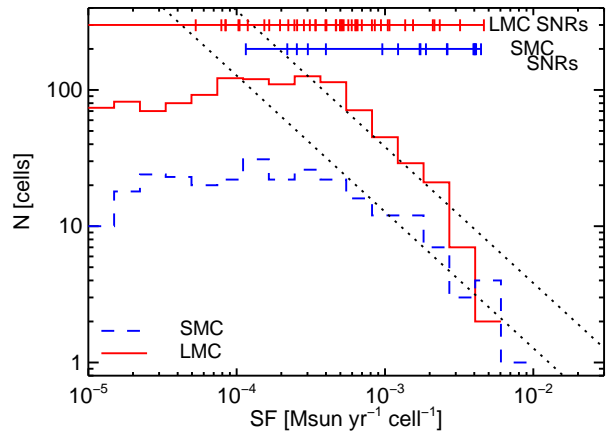}
\caption{Binned histogram of star formation rates averaged over the last 35 Myr in the LMC (red, solid plot) and the SMC
  (blue, dashed plot). The dotted lines are power laws of index $-1$ that intersect the data at a SFR of
  $2\times10^{-3}\,\rm{M_{\odot}\,yr^{-1}\,cell^{-1}}$. The SFR values for the cells that contain the SNRs in both
  galaxies are represented by tick marks on the horizontal rulers.}
\label{SFHist}
\end{figure}

Figure~\ref{SFHist} shows that, above a value of $\sim (2-4)\times 10^{-4} M_\odot ~{\rm yr}^{-1} ~{\rm cell}^{-1}$, the
SFR is indeed distributed as a power law of index $-1$, over more than one order of magnitude in SFR, for both the LMC and
the SMC. To meaningfully compare the SFRs in cells with different physical areas, we have counted the few $24'\times
24'$ cells in the LMC as four $12'\times 12'$ cells with the same SFR, and we have scaled down the SFRs of all the
$12'\times 12'$ cells in the SMC by a factor of 1.44, to account for the fact that this galaxy is 20\% more distant than
the LMC. 

\subsection{Schmidt Law plus star formation rates from H$\alpha$ emission}

\begin{figure}
\includegraphics[width=0.32\textwidth, angle=90]{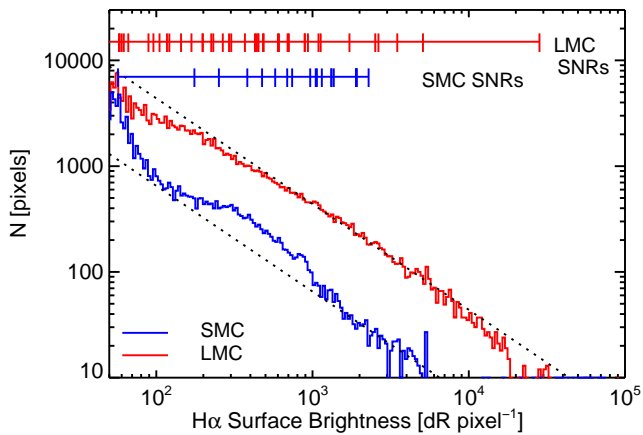}
\caption{Binned histogram of H$\alpha$ surface brightness in the LMC (red) and the SMC (blue). Dotted lines are power laws of
  index $-1$ that intersects the data at $2\times10^{3}$ dR pixel$^{-1}$. The flux values for the cells that contain the
  SNRs in both galaxies are represented by tick marks on the horizontal rulers.}
\label{HaHist}
\end{figure}

Rather than measuring SFR via the resolved stellar populations of the MCs, we can trace it by means of H$\alpha$
emission, which gives us a third probe of gas density. H$\alpha$ is principally powered by photoionisation from O-type
stars, whose numbers are proportional to the SFR. The Schmidt law again connects this SFR to the gas mass column. We
have examined the continuum-subtracted H$\alpha$ emission maps of the LMC and SMC from the SHASSA survey of
\citet{gaustad01:SHASSA} \footnote{The maps are available at \url{http://amundsen.swarthmore.edu/}}.
Figure~\ref{HaHist} shows that both the LMC and the SMC have distributions of H$\alpha$ surface brightness that follow
power laws of index $\sim -1$, this time over 2 orders of magnitude for the SMC, and almost 3 for the LMC. As before, through the
Schmidt law, this implies a power law with index $-1$ for the gas density distribution over 1-2 orders of magnitude in
density. Surface brightness is independent of distance for such nearby galaxies, and can therefore be compared directly
between the LMC and SMC.
 
\subsection{The Interstellar Medium of the Magellanic Clouds}

The density tracers that we have examined here are not free of biases, but they should provide a fairly accurate picture
of the density distribution in the interstellar medium of the MCs. Specific regions within the Clouds, like the LMC bar
or 30 Dor, will probably have density distributions that are different from the simple powerlaws we have found here; our
analysis only applies to the bulk, statistical properties of the gas on galactic scales. Several theoretical studies
have adressed the probability distribution of densities in the interstellar medium: see
\cite{padoan97:IMF_universality}, \cite{passot98:density_distribution_ISM},
\cite{scalo98:interstellar_gas_density_distribution}, and \cite{wada01:multiphase_ISM}. The consensus of these studies
is that, despite local inhomogeneities and deviations, a combination of galaxy-wide and local mechanisms conspire to
produce a lognormal distribution of densities in the gas for a wide range of conditions. At the higher densities where
most stars are formed and most SNe explode, the tail of the lognormal distribution is well reproduced by a power law
with an index $\sim -1$ \citep[see figure 19 in][]{wada07:density_structure_ISM}. This is consistent with our findings
for the MCs.

\section{Discussion}
\label{sec:disc}

The derivation in \S~\ref{physics} ignores many subtleties in what is clearly a complex problem, involving different
physical processes and temporal scales. In this context, the fact that the distribution of gas densities in the MCs
seems to behave like a power law with an index $\sim-1$ does not \textit{prove} that our proposed picture for the
evolution of SNRs is correct. However, our simple (but physically motivated) scenario appears to explain the available
observations quite well, and is consistent with what we know about the bulk properties of the interstellar gas in the
MCs. Furthermore, there are indirect ways to test that the basic scenario is sound.

In Paper~II, we use the compilation of SNR observations from Table 1 and the SFH maps of \citet{harris04:SMC_SFH} and
\citet{harris09:LMC_SFH} to derive the SN rate and delay time distribution in the Magellanic Clouds. Some key results
from that exercise, which we describe briefly here, serve as plausibility arguments for our picture of SNR evolution. A
basic ingredient in the derivation of SN rates is the ``visibility time'' of SNRs (i.e., the time during which a SNR
would be visible, if it were there). We have identified this time with the transition to the radiative phase in
\S~\ref{physics}, but without assigning any specific value to it. This cannot be done using theoretical arguments alone,
because many of the necessary parameters are not known with sufficient accuracy, like the distribution of SN kinetic
energies or the normalization of the cooling function (see Eq.~\ref{rmaxwithepsilon}). The visibility time cannot be
calibrated using individual objects either, because the only SNRs that have reliable ages from light echoes or
historical observations are much younger than the old SNRs in the Sedov stage that form the bulk of our sample. 

In Paper~II, this conundrum is solved by imposing a condition on the delay time distribution: that the vast majority of
stars more massive than $8\,\mathrm{M_{\odot}}$ explode as core collapse SNe, with very few ``silent'' events that
collapse directly to form a black hole without much ejection of material and hence without leaving a SNR behind. If this
condition holds, an absolute value for the SNR visibility time can be obtained by equating the time-integrated SN rate
in the temporal bin of the delay time distribution that is associated with core collapse SNe (in the binning used in
Paper~II, that is all SNe with delays shorter than 35 Myr) to the number of massive stars per total stellar mass
formed. This procedure yields SNR visibility times that depend on the tracer used to determine the local density, and
vary between $13.3$ kyr (for the HI density tracer) and $22.5$ kyr (for the SFR-based density tracer -- see table 2 in
Paper~II). These ages are for regions where the local density corresponds to the mean value of the tracer, $1.5\times
10^{21}~{\rm cm}^{-2}$ and $3.3\times10^{-4}~M_\odot ~{\rm yr}^{-1}\,\mathrm{cell}^{-1}$, respectively (see Figures
\ref{HIHist} and \ref{SFHist}); for values 10 times lower than the mean, the visibility times could be a factor 3 to 4
longer (see Paper~II for details). This is in good agreement with the maximum ages of SNR shells inferred by SNR-pulsar
associations in the Milky Way, which are $\sim60$ kyr \citep{frail94:SNR_lifetime,gaensler95:PSR_SNR_Connection_II}.

Another key result from Paper~II is the total rate of SNe (CC and Type Ia) in the Magellanic Clouds. The rate depends
again on the density tracer used to estimate the visibility time, and varies between
$4.1\pm0.9\times10^{-3}\,\mathrm{SN\,yr^{-1}}$ (for HI) and $2.4\pm0.4\times10^{-3}\,\mathrm{SN\,yr^{-1}}$ (using SFR as
the tracer). These rates, and the visibility times quoted in the previous paragraph, are derived assuming that our SNR
sample is complete and our evolution scenario is correct. As explained in Paper~II, the rates, when normalized per unit
stellar mass, are typical of dwarf irregular galaxies, of which the MCs are prototypes.  The rates are also in loose
agreement with the historical records, which show 2 SNe in the MCs in the last 400 yr \citep[i.e., roughly
$5\times10^{-3}\,\mathrm{SN\,yr^{-1}}$, see discussion in \S~5.3 of][]{badenes08:0509}, although this number of events
is too small to put strong constraints on the SN rate. Adopting the assumption that most SNRs are actually in free
expansion and not in the Sedov stage, as has been invoked by a number of authors to explain the observed uniform size
distribution (see \S~\ref{distribution}), would lead to inconsistencies. Undecelerated SN ejecta reach velocities in
excess of 10000~km~s$^{-1}$, so we would expect the shock velocities in freely expanding SNRs to be in the range $\sim
5000$ to 10000~km~s$^{-1}$. At these velocities, the cutoff radius of $r\sim 30$~pc would translate to maximum ages of 3
to 6 kyr. The 77 SNRs in the MCs would then imply a SN explosion in the MCs once every 40 to 80 years, on average, or
even more frequently if our SNR sample is incomplete. Finally, if the visibility time were fixed at 6~kyr, for instance,
this would correspond to a CC-SN yield (i.e., the number of CC SNe per stellar mass formed) of
$0.022~\mathrm{M_\odot^{-1}}$ (see Paper~II). To produce so many CC-SNe all stars with initial mass above
$3.7~\mathrm{M_\odot}$ in a standard IMF would have to explode, which is in direct contradiction to stellar evolution
theory, and to the semi-empirical initial-final mass relation for WDs
\citep{catalan08:initial_final_mass_WDs,salaris09:WD_initial_mass_function,williams09:lower_mass_SN_progenitors}.

\section{Summary and Conclusions}
\label{sec:summary}

In this paper, we have proposed a physically motivated scenario for the evolution of SNRs in the interstellar medium of
galaxies, and we have applied it to the sample of SNRs in the Magellanic Clouds. We compiled multi-wavelength
observations of the 77 known SNRs in the MCs from the existing literature (Table 1 and Figure \ref{fig-HIMaps}). We
verified that this compilation is fairly complete, and that the size distribution of SNRs is approximately flat, within
the allowed uncertainties, up to a cutoff at $r\sim30$ pc, as discussed by other authors before. We noted that these
features are also present in the larger SNR sample assembled by \cite{long10:M33_SNRs} for M33. In our model, the flat
size distribution can be explained if most SNRs are in the Sedov, decelerating, stage of their expansion, quickly fading
below detection as soon as they reach the radiative stage at an age (size) that depends on the local density. Under
these circumstances, a flat distribution of SNR sizes arises naturally if the probability distribution of densities in
the gas follows a power law with index $-1$. To test this hypothesis, we have examined the distributions of three
different density tracers in the Clouds: HI column density, H$\alpha$ flux, and SFR obtained from resolved stellar
populations. We have verified that these tracers do indeed indicate a density distribution that follows a power law with
index $-1$, over a wide range of densities, and that this agrees with our theoretical understanding of the dynamics of
the interstellar medium. Our scenario implies that SNRs will remain visible for different times at different locations
in the Magellanic Clouds, depending on the local density. This visibility time is a key ingredient in the calculation of
SN rates and SN delay time distributions, which we derive for the Magellanic Clouds in Paper~II. The absolute value of
the visibility time cannot be determined from theoretical arguments alone, but we have used the results from Paper~II to
verify that the range of values we obtain ($13$ to $23$ kyr for regions of mean density, depending on the tracer) is
consistent with the existing limits for the maximum ages of SNRs, and with basic stellar evolution theory.

It would be interesting to extend our analysis of SNR sizes and interstellar medium densities to other galaxies in the
Local Group. On the one hand, this would allow us to test the validity of our SNR evolution scenario in different
settings. On the other hand, the techniques developed here and in Paper~II would yield more accurate SN rates and more
detailed delay time distributions with larger SNR samples, provided the SFHs in the host galaxies could be determined
with enough spatial and temporal resolution. A growing number of nearby galaxies have published SNR catalogues, but
samples of the quality of the one we have assembled here for the Magellanic Clouds or the one published by
\cite{long10:M33_SNRs} for M33 are hard to obtain. Interstellar extinction and distance uncertainties plague the Milky
Way SNRs, and the radio detection limits become comparable to the fluxes of the fainter Magellanic Cloud SNRs for
distances beyond several hundred kpc \citep[see][]{chomiuk09:SNR_Luminosity_Function}. With a judicious investment of
observing time, however, reasonably complete multi-wavelength catalogues of SNRs in a few well-suited nearby galaxies
could become available in the near future.
  
\section*{Acknowledgments}

We are grateful to Jos\'e Luis Prieto, who first suggested using resolved stellar populations around SNRs to constrain
the properties of SN progenitors.  We want to thank the referee for a careful reading of the manuscript, and Dennis
Zaritsky for many helpful discussions about several details of the SFH maps of the MCs. We acknowledge useful
interactions with Laura Chomiuk, Bryan Gaensler, Avishay Gal-Yam, Dave Green, Jack Hughes, Amiel Sternberg, Marten van
Kerkwijk, Jacco Vink, and Eli Waxman. C.B. thanks the Benoziyo Center for Astrophysics for support at the Weizmann
Institute of Science.  D.M. acknowledges support by the Israel Science Foundation and by the DFG through German-Israeli
Project Cooperation grant STE1869/1$-$1.GE625/15$-$1. This research has made use of NASA's Astrophysics Data System
(ADS) Bibliographic Services, as well as the NASA/IPAC Extragalactic Database (NED).


\begin{thebibliography}{90}
\expandafter\ifx\csname natexlab\endcsname\relax\def\natexlab#1{#1}\fi

\bibitem[{{Alves}(2004)}]{alves04:LMC_Distance}
{Alves} D.~R., 2004, New Astronomy Review, 48, 659

\bibitem[{{Badenes}(2010)}]{badenes10:PNAS_review}
{Badenes} C., 2010, Proceedings of the National Academy of Science, 107, 7141 

\bibitem[{{Badenes} {et~al.}(2009){Badenes}, {Harris}, {Zaritsky}, \&
  {Prieto}}]{badenes09:SNRs_LMC}
{Badenes} C., {Harris} J., {Zaritsky} D., {Prieto} J.~L., 2009, \apj, 700, 727

\bibitem[{{Badenes} {et~al.}(2007){Badenes}, {Hughes}, {Bravo}, \&
  {Langer}}]{badenes07:outflows}
{Badenes} C., {Hughes} J.~P., {Bravo} E., {Langer} N., 2007, \apj, 662, 472

\bibitem[{{Badenes} {et~al.}(2008){Badenes}, {Hughes}, {Cassam-Chena{\"i}}, \&
  {Bravo}}]{badenes08:0509}
{Badenes} C., {Hughes} J.~P., {Cassam-Chena{\"i}} G., {Bravo} E., 2008, \apj,
  680, 1149

\bibitem[{{Bamba} {et~al.}(2006){Bamba}, {Ueno}, {Nakajima}, {Mori}, \&
  {Koyama}}]{bamba06:DEM_L241}
{Bamba} A., {Ueno} M., {Nakajima} H., {Mori} K., {Koyama} K., 2006, \aap, 450,
  585

\bibitem[{{Bandiera} \& {Petruk}(2010)}]{bandiera10:Statistics_SNRs}
{Bandiera} R., {Petruk} O., 2010, \aap, 509, A34+

\bibitem[{{Berkhuijsen}(1987)}]{berkhuijsen87:SNRs_N_D}
{Berkhuijsen} E.~M., 1987, \aap, 181, 398

\bibitem[{{Bigiel} {et~al.}(2008){Bigiel}, {Leroy}, {Walter}, {Brinks}, {de
  Blok}, {Madore}, \& {Thornley}}]{bigiel08:SF_law_sub_kpc}
{Bigiel} F., {Leroy} A., {Walter} F., {Brinks} E., {de Blok} W.~J.~G., {Madore}
  B., {Thornley} M.~D., 2008, \aj, 136, 2846

\bibitem[{{Blondin} {et~al.}(1998){Blondin}, {Wright}, {Borkowski}, \&
  {Reynolds}}]{blondin98:Radiative_SNRs}
{Blondin} J.~M., {Wright} E.~B., {Borkowski} K.~J., {Reynolds} S.~P., 1998,
  \apj, 500, 342

\bibitem[{{Boji{\v c}i{\'c}} {et~al.}(2007){Boji{\v c}i{\'c}}, {Filipovi{\'c}},
  {Parker}, {Payne}, {Jones}, {Reid}, {Kawamura}, \& {Fukui}}]{bojicic07:B0513}
{Boji{\v c}i{\'c}} I.~S., {Filipovi{\'c}} M.~D., {Parker} Q.~A., {Payne} J.~L.,
  {Jones} P.~A., {Reid} W., {Kawamura} A., {Fukui} Y., 2007, \mnras, 378, 1237

\bibitem[{{Borkowski} {et~al.}(2006){Borkowski}, {Hendrick}, \&
  {Reynolds}}]{borkowski06:DEML238_DEML249}
{Borkowski} K.~J., {Hendrick} S.~P., {Reynolds} S.~P., 2006, \apj, 652, 1259

\bibitem[{{Catal{\'a}n} {et~al.}(2008){Catal{\'a}n}, {Isern},
  {Garc{\'{\i}}a-Berro}, \& {Ribas}}]{catalan08:initial_final_mass_WDs}
{Catal{\'a}n} S., {Isern} J., {Garc{\'{\i}}a-Berro} E., {Ribas} I., 2008,
  \mnras, 387, 1693

\bibitem[{{Chen} {et~al.}(2006){Chen}, {Wang}, {Gotthelf}, {Jiang}, {Chu}, \&
  {Gruendl}}]{chen06:N157B}
{Chen} Y., {Wang} Q.~D., {Gotthelf} E.~V., {Jiang} B., {Chu} Y., {Gruendl} R.,
  2006, \apj, 651, 237

\bibitem[{Chevalier(1982)}]{chevalier82:selfsimilar}
Chevalier R., 1982, ApJ, 258, 790

\bibitem[{{Chomiuk} \& {Wilcots}(2009)}]{chomiuk09:SNR_Luminosity_Function}
{Chomiuk} L., {Wilcots} E.~M., 2009, \apj, 703, 370

\bibitem[{{Chu} \& {Kennicutt}(1988)}]{chu88:LMC_SNRs_environments}
{Chu} Y.-H., {Kennicutt} Jr. R.~C., 1988, \aj, 96, 1874

\bibitem[{{Cioffi} {et~al.}(1988){Cioffi}, {McKee}, \&
  {Bertschinger}}]{cioffi88:Radiative_SNRs}
{Cioffi} D.~F., {McKee} C.~F., {Bertschinger} E., 1988, \apj, 334, 252

\bibitem[{{Cox}(2005)}]{cox05:ISM}
{Cox} D.~P., 2005, ARA\&A, 43, 337

\bibitem[{{Dopita} {et~al.}(2010){Dopita}, {Blair}, {Long}, {Mutchler},
  {Whitmore}, {Kuntz}, {Balick}, {Bond}, {Calzetti}, {Carollo}, {Disney},
  {Frogel}, {O'Connell}, {Hall}, {Holtzman}, {Kimble}, {MacKenty}, {McCarthy},
  {Paresce}, {Saha}, {Silk}, {Sirianni}, {Trauger}, {Walker}, {Windhorst}, \&
  {Young}}]{dopita10:M83_SNRs}
{Dopita} M.~A., et al., 2010, \apj, 710, 964

\bibitem[{{Dwarkadas}(2005)}]{dwarkadas05:SNR-Bubbles_1D}
{Dwarkadas} V.~V., 2005, ApJ, 630, 892

\bibitem[{{Dwarkadas}(2007)}]{dwarkadas07:SNRs_Bubbles_WR}
---, 2007, \apj, 667, 226

\bibitem[{{Ferri{\`e}re}(2001)}]{ferriere01:ISM}
{Ferri{\`e}re} K.~M., 2001, Rev.\ Mod.\ Phys., 73, 1031

\bibitem[{{Fesen} {et~al.}(1985){Fesen}, {Blair}, \&
  {Kirshner}}]{fesen85:Evolved_SNRs}
{Fesen} R.~A., {Blair} W.~P., {Kirshner} R.~P., 1985, \apj, 292, 29

\bibitem[{{Filipovi{\'c}} {et~al.}(2002){Filipovi{\'c}}, {Bohlsen}, {Reid},
  {Staveley-Smith}, {Jones}, {Nohejl}, \&
  {Goldstein}}]{filipovic02:SMC_radio_4bands}
{Filipovi{\'c}} M.~D., {Bohlsen} T., {Reid} W., {Staveley-Smith} L., {Jones}
  P.~A., {Nohejl} K., {Goldstein} G., 2002, \mnras, 335, 1085

\bibitem[{{Filipovic} {et~al.}(1998){Filipovic}, {Haynes}, {White}, \&
  {Jones}}]{filipovic98:MC_Radio_Discrete}
{Filipovic} M.~D., {Haynes} R.~F., {White} G.~L., {Jones} P.~A., 1998, \aaps,
  130, 421

\bibitem[{{Filipovic} {et~al.}(1995){Filipovic}, {Haynes}, {White}, {Jones},
  {Klein}, \& {Wielebinski}}]{filipovic95:LMC_Radio_5Bands}
{Filipovic} M.~D., {Haynes} R.~F., {White} G.~L., {Jones} P.~A., {Klein} U.,
  {Wielebinski} R., 1995, \aaps, 111, 311

\bibitem[{{Filipovi{\'c}} {et~al.}(2005){Filipovi{\'c}}, {Payne}, {Reid},
  {Danforth}, {Staveley-Smith}, {Jones}, \& {White}}]{filipovic05:SMc_SNRs}
{Filipovi{\'c}} M.~D., {Payne} J.~L., {Reid} W., {Danforth} C.~W.,
  {Staveley-Smith} L., {Jones} P.~A., {White} G.~L., 2005, \mnras, 364, 217

\bibitem[{{Frail} {et~al.}(1994){Frail}, {Goss}, \&
  {Whiteoak}}]{frail94:SNR_lifetime}
{Frail} D.~A., {Goss} W.~M., {Whiteoak} J.~B.~Z., 1994, \apj, 437, 781

\bibitem[{{Fusco-Femiano} \&
  {Preite-Martinez}(1984)}]{fusco-femiano84:LMC_SNRs}
{Fusco-Femiano} R., {Preite-Martinez} A., 1984, \apj, 281, 593

\bibitem[{{Gaensler} {et~al.}(2003){Gaensler}, {Hendrick}, {Reynolds}, \&
  {Borkowski}}]{gaensler03:0453}
{Gaensler} B.~M., {Hendrick} S.~P., {Reynolds} S.~P., {Borkowski} K.~J., 2003,
  \apjl, 594, L111

\bibitem[{{Gaensler} \& {Johnston}(1995)}]{gaensler95:PSR_SNR_Connection_II}
{Gaensler} B.~M., {Johnston} S., 1995, \mnras, 277, 1243

\bibitem[{{Gaetz} {et~al.}(2000){Gaetz}, {Butt}, {Edgar}, {Eriksen},
  {Plucinsky}, {Schlegel}, \& {Smith}}]{gaetz00:E0102}
{Gaetz} T.~J., {Butt} Y.~M., {Edgar} R.~J., {Eriksen} K.~A., {Plucinsky} P.~P.,
  {Schlegel} E.~M., {Smith} R.~K., 2000, \apjl, 534, L47

\bibitem[{{Gaustad} {et~al.}(2001){Gaustad}, {McCullough}, {Rosing}, \& {Van
  Buren}}]{gaustad01:SHASSA}
{Gaustad} J.~E., {McCullough} P.~R., {Rosing} W., {Van Buren} D., 2001, \pasp,
  113, 1326

\bibitem[{{Green}(1984)}]{green84:SNR_Statistics}
{Green} D.~A., 1984, \mnras, 209, 449

\bibitem[{{Harris} \& {Zaritsky}(2001)}]{harris01:StarFISH}
{Harris} J., {Zaritsky} D., 2001, \apjs, 136, 25

\bibitem[{{Harris} \& {Zaritsky}(2004)}]{harris04:SMC_SFH}
---, 2004, \aj, 127, 1531

\bibitem[{{Harris} \& {Zaritsky}(2009)}]{harris09:LMC_SFH}
---, 2009, \aj, 138, 1243

\bibitem[{{Hatzidimitriou} \& {Hawkins}(1989)}]{hatzidimitriou89:SMC_structure}
{Hatzidimitriou} D., {Hawkins} M.~R.~S., 1989, \mnras, 241, 667

\bibitem[{Hendrick {et~al.}(2003)Hendrick, Borkowski, \&
  Reynolds}]{hendrick03:LMC_SNRs}
Hendrick S., Borkowski K., Reynolds S.~P., 2003, ApJ, 593, 370

\bibitem[{{Hendrick} {et~al.}(2005){Hendrick}, {Reynolds}, \&
  {Borkowski}}]{hendrick05:B0049}
{Hendrick} S.~P., {Reynolds} S.~P., {Borkowski} K.~J., 2005, \apjl, 622, L117

\bibitem[{{Hilditch} {et~al.}(2005){Hilditch}, {Howarth}, \&
  {Harries}}]{hilditch05:SMC_Distance}
{Hilditch} R.~W., {Howarth} I.~D., {Harries} T.~J., 2005, \mnras, 357, 304

\bibitem[{{Hughes} {et~al.}(2007){Hughes}, {Staveley-Smith}, {Kim}, {Wolleben},
  \& {Filipovi{\'c}}}]{hughes07:LMC_21cm_ATCA}
{Hughes} A., {Staveley-Smith} L., {Kim} S., {Wolleben} M., {Filipovi{\'c}} M.,
  2007, \mnras, 382, 543

\bibitem[{{Hughes} {et~al.}(1984){Hughes}, {Helfand}, \&
  {Kahn}}]{hughes84:SNR_N_D}
{Hughes} J.~P., {Helfand} D.~J., {Kahn} S.~M., 1984, \apjl, 281, L25

\bibitem[{{Hughes} {et~al.}(2006){Hughes}, {Rafelski}, {Warren}, {Rakowski},
  {Slane}, {Burrows}, \& {Nousek}}]{hughes06:N23}
{Hughes} J.~P., {Rafelski} M., {Warren} J.~S., {Rakowski} C., {Slane} P.,
  {Burrows} D., {Nousek} J., 2006, \apjl, 645, L117

\bibitem[{{Hwang} {et~al.}(2001){Hwang}, {Petre}, {Holt}, \&
  {Szymkowiak}}]{hwang01:N158A}
{Hwang} U., {Petre} R., {Holt} S.~S., {Szymkowiak} A.~E., 2001, \apj, 560, 742

\bibitem[{{Kennicutt}(1989)}]{kennicutt89:SFR_galdisks}
{Kennicutt} Jr. R.~C., 1989, \apj, 344, 685

\bibitem[{{Kennicutt}(1998)}]{kennicutt98:SF_review}
---, 1998, \araa, 36, 189

\bibitem[{{Kim} {et~al.}(1998){Kim}, {Chu}, {Staveley-Smith}, \&
  {Smith}}]{kim98:N44}
{Kim} S., {Chu} Y., {Staveley-Smith} L., {Smith} R.~C., 1998, \apj, 503, 729

\bibitem[{{Kim} {et~al.}(2007){Kim}, {Rosolowsky}, {Lee}, {Kim}, {Jung},
  {Dopita}, {Elmegreen}, {Freeman}, {Sault}, {Kesteven}, {McConnell}, \&
  {Chu}}]{kim07:HI_Clouds_LMC}
{Kim} S., {Rosolowsky} E., {Lee} Y., {Kim} Y., {Jung} Y.~C., {Dopita} M.~A.,
  {Elmegreen} B.~G., {Freeman} K.~C., {Sault} R.~J., {Kesteven} M., {McConnell}
  D., {Chu} Y., 2007, \apjs, 171, 419

\bibitem[{{Kim} {et~al.}(2003){Kim}, {Staveley-Smith}, {Dopita}, {Sault},
  {Freeman}, {Lee}, \& {Chu}}]{kim03:LMC_HI_Parkes_ATCA}
{Kim} S., {Staveley-Smith} L., {Dopita} M.~A., {Sault} R.~J., {Freeman} K.~C.,
  {Lee} Y., {Chu} Y., 2003, \apjs, 148, 473

\bibitem[{{Long} {et~al.}(2010){Long}, {Blair}, {Winkler}, {Becker}, {Gaetz},
  {Ghavamian}, {Helfand}, {Hughes}, {Kirshner}, {Kuntz}, {McNeil}, {Pannuti},
  {Plucinsky}, {Saul}, {T{\"u}llmann}, \& {Williams}}]{long10:M33_SNRs}
{Long} K.~S., et al., 2010, \apjs, 187, 495

\bibitem[{{Mac Low} \& {McCray}(1988)}]{maclow88:superbubbles}
{Mac Low} M., {McCray} R., 1988, \apj, 324, 776

\bibitem[{{Magnier} {et~al.}(1997){Magnier}, {Primini}, {Prins}, {van
  Paradijs}, \& {Lewin}}]{magnier97:M31_SNRs_ROSAT}
{Magnier} E.~A., {Primini} F.~A., {Prins} S., {van Paradijs} J., {Lewin}
  W.~H.~G., 1997, \apj, 490, 649

\bibitem[{{Maoz} \& {Rix}(1993)}]{maoz93:grav_lensing_statistics}
{Maoz} D., {Rix} H., 1993, \apj, 416, 425

\bibitem[{Mathewson {et~al.}(1984)Mathewson, Ford, Dopita, Tuohy, Mills, \&
  Turtle}]{mathewson84:SNR-Magellanic-Clouds}
Mathewson D., Ford V., Dopita M., Tuohy I., Mills B., Turtle A., 1984, ApJS,
  55, 189

\bibitem[{{McKee} \& {Ostriker}(1977)}]{mckee77:ISM_Theory}
{McKee} C.~F., {Ostriker} J.~P., 1977, \apj, 218, 148

\bibitem[{{Mills} {et~al.}(1984){Mills}, {Turtle}, {Little}, \&
  {Durdin}}]{mills84:Radio_SNRs}
{Mills} B.~Y., {Turtle} A.~J., {Little} A.~G., {Durdin} J.~M., 1984, Australian
  Journal of Physics, 37, 321

\bibitem[Murphy Williams et al.(2010)]{2010AAS...21545416M} Murphy 
Williams, R.~N., Dickel, J.~R., Chu, Y., Points, S., Winkler, F., Johnson, 
M., \& Lodder, K.\ 2010, Bulletin of the American Astronomical Society, 41, 470 

\bibitem[{{Ng} {et~al.}(2008){Ng}, {Gaensler}, {Staveley-Smith}, {Manchester},
  {Kesteven}, {Ball}, \& {Tzioumis}}]{ng08:SN1987A}
{Ng} C.-Y., {Gaensler} B.~M., {Staveley-Smith} L., {Manchester} R.~N.,
  {Kesteven} M.~J., {Ball} L., {Tzioumis} A.~K., 2008, \apj, 684, 481

\bibitem[{{Oey} \& {Clarke}(1997)}]{oey97:superbubble_sizes}
{Oey} M.~S., {Clarke} C.~J., 1997, \mnras, 289, 570

\bibitem[{{Padoan} {et~al.}(1997){Padoan}, {Nordlund}, \&
  {Jones}}]{padoan97:IMF_universality}
{Padoan} P., {Nordlund} A., {Jones} B.~J.~T., 1997, \mnras, 288, 145

\bibitem[{{Park} {et~al.}(2003{\natexlab{a}}){Park}, {Burrows}, {Garmire},
  {Nousek}, {Hughes}, \& {Williams}}]{park03:N49}
{Park} S., {Burrows} D.~N., {Garmire} G.~P., {Nousek} J.~A., {Hughes} J.~P.,
  {Williams} R.~M., 2003{\natexlab{a}}, \apj, 586, 210

\bibitem[{{Park} {et~al.}(2003{\natexlab{b}}){Park}, {Hughes}, {Burrows},
  {Slane}, {Nousek}, \& {Garmire}}]{park03:E0103}
{Park} S., {Hughes} J.~P., {Burrows} D.~N., {Slane} P.~O., {Nousek} J.~A.,
  {Garmire} G.~P., 2003{\natexlab{b}}, \apjl, 598, L95

\bibitem[{{Park} {et~al.}(2003{\natexlab{c}}){Park}, {Hughes}, {Slane},
  {Burrows}, {Warren}, {Garmire}, \& {Nousek}}]{park03:N49B}
{Park} S., {Hughes} J.~P., {Slane} P.~O., {Burrows} D.~N., {Warren} J.~S.,
  {Garmire} G.~P., {Nousek} J.~A., 2003{\natexlab{c}}, \apjl, 592, L41

\bibitem[{{Passot} \&
  {V{\'a}zquez-Semadeni}(1998)}]{passot98:density_distribution_ISM}
{Passot} T., {V{\'a}zquez-Semadeni} E., 1998, \pre, 58, 4501

\bibitem[{{Payne} {et~al.}(2008){Payne}, {White}, \&
  {Filipovi{\'c}}}]{payne08:LMCSNRs}
{Payne} J.~L., {White} G.~L., {Filipovi{\'c}} M.~D., 2008, \mnras, 383, 1175

\bibitem[{{Plucinsky} {et~al.}(2008){Plucinsky}, {Williams}, {Long}, {Gaetz},
  {Sasaki}, {Pietsch}, {T{\"u}llmann}, {Smith}, {Blair}, {Helfand}, {Hughes},
  {Winkler}, {de Avillez}, {Bianchi}, {Breitschwerdt}, {Edgar}, {Ghavamian},
  {Grindlay}, {Haberl}, {Kirshner}, {Kuntz}, {Mazeh}, {Pannuti}, {Shporer}, \&
  {Thilker}}]{plucinsky08:ChASeM33}
{Plucinsky} P.~P., et al., 2008, \apjs, 174, 366

\bibitem[{{Rakowski} {et~al.}(2006){Rakowski}, {Badenes}, {Gaensler},
  {Gelfand}, {Hughes}, \& {Slane}}]{rakowski05:G337}
{Rakowski} C.~E., {Badenes} C., {Gaensler} B.~M., {Gelfand} J.~D., {Hughes}
  J.~P., {Slane} P.~O., 2006, \apj, 646, 982

\bibitem[{{Reyes-Iturbide} {et~al.}(2008){Reyes-Iturbide}, {Rosado}, \&
  {Vel{\'a}zquez}}]{reyes-iturbide08:N120}
{Reyes-Iturbide} J., {Rosado} M., {Vel{\'a}zquez} P.~F., 2008, \aj, 136, 2011

\bibitem[{{Salaris} {et~al.}(2009){Salaris}, {Serenelli}, {Weiss}, \& {Miller
  Bertolami}}]{salaris09:WD_initial_mass_function}
{Salaris} M., {Serenelli} A., {Weiss} A., {Miller Bertolami} M., 2009, \apj,
  692, 1013

\bibitem[{{Scalo} {et~al.}(1998){Scalo}, {Vazquez-Semadeni}, {Chappell}, \&
  {Passot}}]{scalo98:interstellar_gas_density_distribution}
{Scalo} J., {Vazquez-Semadeni} E., {Chappell} D., {Passot} T., 1998, \apj, 504,
  835

\bibitem[{{Schmidt}(1959)}]{schmidt59:SFR}
{Schmidt} M., 1959, \apj, 129, 243

\bibitem[{{Seward} {et~al.}(2006){Seward}, {Williams}, {Chu}, {Dickel},
  {Smith}, \& {Points}}]{seward06:N9}
{Seward} F.~D., {Williams} R.~M., {Chu} Y., {Dickel} J.~R., {Smith} R.~C.,
  {Points} S.~D., 2006, \apj, 640, 327

\bibitem[{{Smith} {et~al.}(2000){Smith}, {Leiton}, \&
  {Pizarro}}]{smith00:MCELS}
{Smith} C., {Leiton} R., {Pizarro} S., 2000, in Astronomical Society of the
  Pacific Conference Series, Vol. 221, Stars, Gas and Dust in Galaxies:
  Exploring the Links, {D.~Alloin, K.~Olsen, \& G.~Galaz}, ed., pp. 83--+

\bibitem[{{Stanimirovic} {et~al.}(1999){Stanimirovic}, {Staveley-Smith},
  {Dickey}, {Sault}, \& {Snowden}}]{stanimirovic99MNRAS.302..417S}
{Stanimirovic} S., {Staveley-Smith} L., {Dickey} J.~M., {Sault} R.~J.,
  {Snowden} S.~L., 1999, \mnras, 302, 417

\bibitem[{{Subramanian} \& {Subramaniam}(2009)}]{subramanian09:LMC_SMC_Depth}
{Subramanian} S., {Subramaniam} A., 2009, \aap, 496, 399

\bibitem[{{Townsley} {et~al.}(2006){Townsley}, {Broos}, {Feigelson}, {Brandl},
  {Chu}, {Garmire}, \& {Pavlov}}]{townsley06:30Dor}
{Townsley} L.~K., {Broos} P.~S., {Feigelson} E.~D., {Brandl} B.~R., {Chu} Y.,
  {Garmire} G.~P., {Pavlov} G.~G., 2006, \aj, 131, 2140

\bibitem[{Truelove \& McKee(1999)}]{truelove99:adiabatic-SNRs}
Truelove J., McKee C., 1999, ApJS, 120, 299

\bibitem[{van~der Heyden {et~al.}(2004)van~der Heyden, Bleeker, \&
  Kaastra}]{heyden03:SMC_SNRs}
van~der Heyden K., Bleeker J., Kaastra J., 2004, A\&A, 421, 1031 

\bibitem[{{van der Marel} \& {Cioni}(2001)}]{vandermarel01:LMC_inclination}
{van der Marel} R.~P., {Cioni} M., 2001, \aj, 122, 1807

\bibitem[{{Wada} \& {Norman}(2001)}]{wada01:multiphase_ISM}
{Wada} K., {Norman} C.~A., 2001, \apj, 547, 172

\bibitem[{{Wada} \& {Norman}(2007)}]{wada07:density_structure_ISM}
---, 2007, \apj, 660, 276

\bibitem[{{Warren} {et~al.}(2003){Warren}, {Hughes}, \&
  {Slane}}]{warren03:N63A}
{Warren} J.~S., {Hughes} J.~P., {Slane} P.~O., 2003, \apj, 583, 260

\bibitem[{{Williams} {et~al.}(2009){Williams}, {Bolte}, \&
  {Koester}}]{williams09:lower_mass_SN_progenitors}
{Williams} K.~A., {Bolte} M., {Koester} D., 2009, \apj, 693, 355

\bibitem[{{Williams} \& {Chu}(2005)}]{williams05:DEML316}
{Williams} R.~M., {Chu} Y., 2005, \apj, 635, 1077

\bibitem[{{Williams} {et~al.}(1999){Williams}, {Chu}, {Dickel}, {Petre},
  {Smith}, \& {Tavarez}}]{williams99:LMC_SNR_Atlas}
{Williams} R.~M., {Chu} Y.-H., {Dickel} J.~R., {Petre} R., {Smith} R.~C.,
  {Tavarez} M., 1999, \apjs, 123, 467

\bibitem[{{Williams} {et~al.}(2000){Williams}, {Petre}, {Chu}, \&
  {Chen}}]{williams00:0540}
{Williams} R.~M., {Petre} R., {Chu} Y., {Chen} C., 2000, \apjl, 536, L27

\bibitem[{Woltjer(1972)}]{woltjer72:SNR-review}
Woltjer L., 1972, ARA\&A, 10, 129

\bibitem[{{Zaritsky} {et~al.}(2004){Zaritsky}, {Harris}, {Thompson}, \&
  {Grebel}}]{zaritsky04:MCPS}
{Zaritsky} D., {Harris} J., {Thompson} I.~B., {Grebel} E.~K., 2004, \aj, 128,
  1606

\end{thebibliography}

\end{document}